\documentclass{aastex631}

\usepackage{MnSymbol,wasysym}
\usepackage{rotating}
\usepackage{longtable}
\usepackage{soul}
\shorttitle{Exop(lan)ets}
\shortauthors{Astrobites Collaboration}
\graphicspath{{./}{figures/}}
\usepackage{blindtext, rotating}

\begin{document}

\title{First Detections of Exop(lan)ets: Observations and Follow-Ups of the Floofiest Transits on Zoom}

\author[0000-0002-6650-3829]{Sabina Sagynbayeva}
\affiliation{Astrobites Collaboration, American Astronomical Society, Washington, D.C. 20006, USA}
\affiliation{Department of Physics and Astronomy, Stony Brook University, Stony Brook, NY 11794, USA}
\author[0000-0002-8984-4319]{Briley L. Lewis}
\affiliation{Department of Physics and Astronomy, UCLA, Los Angeles, CA 90024, USA}
\affiliation{Astrobites Collaboration, American Astronomical Society, Washington, D.C. 20006, USA}

\author[0000-0002-4219-6908]{Graham M. Doskoch}
\affiliation{Department of Physics and Astronomy, West Virginia University, P.O. Box 6315, Morgantown, WV 26506, USA}
\affiliation{Center for Gravitational Waves and Cosmology, West Virginia University, Chestnut Ridge Research Building, Morgantown, WV 26505, USA}
\affiliation{Astrobites Collaboration, American Astronomical Society, Washington, D.C. 20006, USA}

\author[0000-0003-4310-3440]{Ali Crisp}
\affiliation{Department of Physics and Astronomy, Louisiana State University, Baton Rouge, LA 70803}
\affiliation{Astrobites Collaboration, American Astronomical Society, Washington, D.C. 20006, USA}

\author[0000-0002-2361-5812]{Catherine A. Clark}
\affiliation{Northern Arizona University, 527 South Beaver Street, Flagstaff, AZ 86011, USA}
\affiliation{Lowell Observatory, 1400 West Mars Hill Road, Flagstaff, AZ 86001, USA}
\affiliation{Astrobites Collaboration, American Astronomical Society, Washington, D.C. 20006, USA}

\author[0000-0003-2294-4187]{Katya Gozman}
\affiliation{Department of Astronomy, University of Michigan, 1085 S. University Ave, Ann Arbor, MI 48109-1107, USA}
\affiliation{Astrobites Collaboration, American Astronomical Society, Washington, D.C. 20006, USA}
\author[0000-0002-3475-7648]{Gourav Khullar}
\affiliation{Astrobites Collaboration, American Astronomical Society, Washington, D.C. 20006, USA}
\affiliation{Department of Astronomy and Astrophysics, University of
Chicago, 5640 South Ellis Avenue, Chicago, IL 60637}
\affiliation{Kavli Institute for Cosmological Physics, University of
Chicago, 5640 South Ellis Avenue, Chicago, IL 60637}
\affiliation{Kavli Institute for Astrophysics \& Space Research, Massachusetts Institute of Technology, Cambridge, MA 02139, USA}

\author[0000-0001-9678-0299]{Haley Wahl}
\affiliation{Department of Physics and Astronomy, West Virginia University, P.O. Box 6315, Morgantown, WV 26506, USA}
\affiliation{Center for Gravitational Waves and Cosmology, West Virginia University, Chestnut Ridge Research Building, Morgantown, WV 26505, USA}
\affiliation{Astrobites Collaboration, American Astronomical Society, Washington, D.C. 20006, USA}

\author[0000-0002-0786-7307]{Jenny K. Calahan}
\affiliation{University of Michigan, 323 West Hall, 1085 South University Avenue, Ann Arbor, MI 48109, USA }
\affiliation{Astrobites Collaboration, American Astronomical Society, Washington, D.C. 20006, USA}

%\author{Lina Kimmig}
%\affiliation{tbd}

\author[0000-0001-9482-7794]{Mark Popinchalk}
\affiliation{Department of Astrophysics, American Museum of Natural History, Central Park West at 79th Street, New York, NY 10034, USA}
\affiliation{Physics, The Graduate Center, City University of New York, New York, NY 10016, USA}
\affiliation{Department of Physics and Astronomy, Hunter College, City University of New York, 695 Park Avenue, New York, NY 10065, USA}
\affiliation{Astrobites Collaboration, American Astronomical Society, Washington, D.C. 20006, USA}

\author{Samuel Factor}
\affiliation{Department of Astronomy, The University of Texas at Austin, 2515 Speedway Boulevard Stop C1400, Austin, TX 78712, USA}
\affiliation{Astrobites Collaboration, American Astronomical Society, Washington, D.C. 20006, USA}

\author[0000-0003-4591-3201]{Macy Huston}
\affiliation{Department of Astronomy \& Astrophysics, The Pennsylvania State University, University Park, PA 16802, USA}
\affiliation{Penn State Extraterrestrial Intelligence Center, The Pennsylvania State University, University Park, PA 16802, USA}
\affiliation{Center for Exoplanets and Habitable Worlds, The Pennsylvania State University, University Park, PA 16802, USA}
\affiliation{Astrobites Collaboration, American Astronomical Society, Washington, D.C. 20006, USA}

\author[0000-0003-0965-605X]{Pratik Gandhi}
\affiliation{Department of Physics and Astronomy, UC Davis, 1 Shields Avenue, Davis, CA 95618, USA}
\affiliation{Astrobites Collaboration, American Astronomical Society, Washington, D.C. 20006, USA}

\author{Isabella Trierweiler}
\affiliation{Department of Physics and Astronomy, UCLA, Los Angeles, CA 90024, USA}
\affiliation{Astrobites Collaboration, American Astronomical Society, Washington, D.C. 20006, USA}

\author[0000-0002-0244-6650]{Suchitra Narayanan}
\affiliation{Institute for Astronomy, University of Hawai`i at Mānoa, 2680 Woodlawn Dr., Honolulu, HI 96822, USA}\affiliation{Astrobites Collaboration, American Astronomical Society, Washington, D.C. 20006, USA}

\author[0000-0002-2072-6541]{Jonathan Brande}
\affiliation{Department of Physics and Astronomy, University of Kansas, 1082 Malott, 1251 Wescoe Hall Dr., Lawrence, KS 66045, USA}
\affiliation{Astrobites Collaboration, American Astronomical Society, Washington, D.C. 20006, USA}

\author[0000-0002-6747-2745]{Michael M. Foley}
\affiliation{Astrobites Collaboration, American Astronomical Society, Washington, D.C. 20006, USA}
\affiliation{Center for Astrophysics $\vert$ Harvard \& Smithsonian, 60 Garden St.,
Cambridge, MA, 02138}

\author[0000-0003-3881-1397]{Olivia R. Cooper}
\affiliation{Astrobites Collaboration, American Astronomical Society, Washington, D.C. 20006, USA}
\affiliation{Department of Astronomy, The University of Texas at Austin, 2515 Speedway Boulevard Stop C1400, Austin, TX 78712, USA}

\author[0000-0002-9544-0118]{Ben Cassese}
\affiliation{Astrobites Collaboration, American Astronomical Society, Washington, D.C. 20006, USA}
\affiliation{Department of Astronomy, Columbia University, 550 W. 120th St, New York City, New York 10027, USA}



\begin{abstract}

With the proliferation of online Zoom\footnote{\url{https://zoom.us/}} meetings as a means of doing science in the 2020s, astronomers have made new and unexpected Target of Opportunity (ToO) observations. Chief among these ToOs are observations of exop(lan)ets, or ``exopets." Building on the work of \citet{Mayorga2021} -- whose work characterized the rotational variations of ``floofy" objects -- we model exopets using methods similar to those used for exoplanetary transits. We present data collected for such exopet Zoom transits through a citizen science program in the month of February 2022. The dataset includes parameters like exopet color, floofiness, transit duration, and percentage of Zoom screen covered during the event. For some targets, we also present microlensing and direct imaging data. Using results from our modelling of 62 exopet observations as transits, microlensing, and direct imaging events, we discuss our inferences of exopet characteristics like their masses, sizes, orbits, colors, and floofiness. 

\end{abstract}



\section{Introduction} \label{sec:intro}

%Sabina:
For more than two years, humanity has been examining new methods of adjusting to work-from-home due to the COVID-19 pandemic. Scientists have tried everything: whipped coffee, sourdough bread, and even questioning whether everything is made out of cake\footnote{The question of whether we live in a giant cake requires further examination.}. One particular adaptation has had an outsized impact and enabled academics to efficiently work from their couches: Zoom. Following rapid widespread adoption of the platform and its contemporaries, Zoom meetings have become the most popular way to get science done. Even a new method of holding conferences - a hybrid method of virtual and in-person meetings - has been introduced, ensuring that the future of science is done via Zoom. Although Zoom meetings are primarily meant to facilitate other scientific research programs, the precision and prevalence of Zoom cameras mean that they themselves have become instruments of discovery in their own right. Over two years of casual observation, we noted occasional drops in the brightness of a Zoom image of our far-flung collaborators, not unlike the dip in brightness cause by an exoplanet during transit \citep{udalski2002, konacki2003}. We attribute these new variations to exop(lan)ets, or ``exopets", defined as pets that inhabit other homes, and began a more systematic survey for them in February 2022.

The great advances in exoplanet science over the last decade have largely been driven by the introduction of new technologies such as the Kepler and TESS missions \citep{borucki2010kepler,guerrero2021tess}. These have resulted in the confirmation of more than 5,000 exoplanets, which in turn has allowed for population-level statistical analyses that illuminated the wide variety of system types and configurations \citep{2017AAS...22914616A, batalha2014exploring}. Similarly, with the standardization of work-from-home Zoom meetings, we are entering a regime where it is possible to perform pup-ulation analysis and cat-egorization of exopets. Although human reactions to Zoom have been mixed \citep{serhan2020transitioning}, most agree that the discovery of exopets has been a highlight of this new work-from-home way of life. %I'm just throwing things out there, take and edit as you want - MP

%%I think we need a transition here
% Maybe leave this paragraph out? It feels like a technical detail that would be best outside the introduction. -BC

% \citet{Mayorga2021} identified rotational variations of floofy objects they discovered using optical observations with 346 frames. They assumed spherical symmetry in their study in order to identify color-magnitude variability. In our work, we do not use the same assumption because this assumption is not necessary when calculating the flux received by a Zoom screen. Instead we assume that a Zoom screen is rectangular, and we compute the length of an exopet based on the transit depth and floofiness reported by the community.

%%probably need a clearer statement of goals for the paper before diving into the detailed summary here?
We, the members of the Astrobites Collaboration \citep{astrobites2017,astrobites2019}\footnote{Astrobites: \href{https://astrobites.org/}{https://astrobites.org/}}, share both the theoretical foundations for exop(lan)et (a.k.a. exopet) transits, the results of our survey, and its possible implications. In Section \ref{sec:theory}, we lay out our transit model and treatment of variations of exopet floofiness. We then describe our data collection methods in Section \ref{sec:obs}. In Section \ref{sec:results}, we summarize our observations of these objects via multiple methods --- transits, microlensing, spectroscopy, and imaging observations for the closest targets --- and the implications of these observations for our physical understanding of the observed systems. Finally, in Sections \ref{sec:disc} and \ref{sec:conc}, we connect our new observations to the broader field of exopet science and discuss future work. Definitions of new terms used in this paper are provided in Table \ref{tab:definitions} in the Appendix. The appendix also contains follow up in-situ observations in Appendix \ref{sec:insitu} and the full data set in Appendix \ref{app-data}.

%^^ Could we add something here like: Fluxes and Magnitudes are calibrated to the AB magnitude system. We assume a $$\Lambda$CDM cosmology, that has graciously facilitated the formation and evolution of exopets. - GK

\section{Exopet Transit Theory}\label{sec:theory}

%Here we define exopet transits and discuss their simulation.
%I don't think this is necessary! We can just jump into the definition. or maybe we don't need to have a subheading for the first bit, that can just be the intro to it

%\subsection{Defining Exopet Transits}\label{sec:transit_theory}
%% Not sure if this will end up being way too transit-centric, so def lmk if y'all think I should cut things - AC
Thanks to NASA's \textit{Kepler} and \textit{TESS} missions, nearly 4,000 of the 5,000+ confirmed exoplanets have been discovered using the transit photometry method \citep{2017AAS...22914616A, borucki2010kepler, guerrero2021tess}.  Similarly, our primary method for exopet detection is their Zoom transits. We define a similar method to exoplanet transits to describe our exopet transits. From our observations, we have known values of transit duration $T_{paw}$, transit depth $\delta$, and exopet floofiness. If we assume that the exopet's transit is at an inclination of 90$^{\circ}$ and an orbital period of 1, we can approximate the exopet radius $L_{paw}$ as follows: 

\begin{equation}\label{pet_radius}
    \delta = \frac{L_{paw}^{2}}{R_{\smiley}^{2}} \Longrightarrow L_{paw} = \sqrt{\delta R_{\smiley}^{2}}
\end{equation}

Our approximation of the host radius, $R_{\smiley}$, is discussed in Section \ref{sec:transits}.

\subsection{Light Curves and Simulations}\label{sec:light-curves-and-simulations}

While it may be difficult to predict the characteristics of exopet distributions, it is certainly possible to simulate individual exopet transits. At any given point in the transit, the exopet can be represented as a projection onto the two-dimensional rectangular image showing the field-of-view of its webcam. We can then define the flux from the webcam as the fraction of its field-of-view not taken up by the projection. By choosing projections of appropriate size and shape, we can simulate the light curve of the webcam during the transit. While exoplanets often appear as circles, exopets produce projections of more complicated shapes. Additionally, these projections change in time as an exopet moves its appendages to propel itself forward. For the purpose of simulating light curves, however, a static projection may be used for the sake of simplicity.

We use an algorithm similar to the one implemented by Sandford \& Kipping in the Python package $\tt EightBitTransit$\footnote{$\tt EightBitTransit$: \href{https://github.com/esandford/EightBitTransit}{https://github.com/esandford/EightBitTransit}} \citep{sandford2019eightbittransit,sandford2019shadow}. We begin with a blank rectangle of width $w$ and height $h$ representing the image formed by a webcam. The rectangle is divided into a large number of discrete blocks of width $\Delta w$ and height $\Delta h$. At each time $t$, the block at location $(x_i,y_j)$ is assigned a quantity $b_{ij}(t)$, which is 1 if part of the exopet lies in the block and 0 otherwise. The normalized flux from the webcam, $f$, is then

\begin{equation}
    f(t)=1-\frac{\text{number of blocks covered}}{\text{total number of blocks}}=1-\frac{\sum_{i=0}^{w/\Delta w}\sum_{j=0}^{h/\Delta h}b_{ij}(t)}{(w/\Delta w)(h/\Delta h)}
\end{equation}\label{eqn:flux}

%%Citations!! - Graham

\begin{figure}
\centering
	\includegraphics[width=0.8\columnwidth]{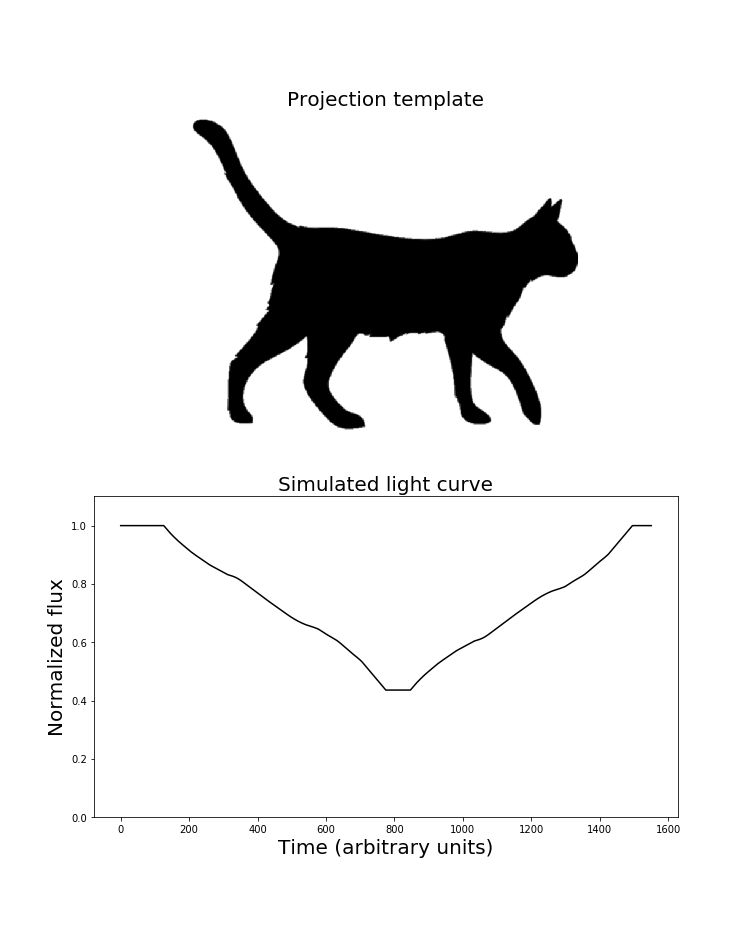}
    \caption{Sample projection template of a cat. Assuming the cat is approximately 1 meter long and walks a distance of 1 meter from a person's webcam, the template produces the light curve shown above. The cat's shape ensures that the flux at the beginning and end of the transit decreases and increases at a reasonably steady rate. Template originally from \url{kindpng.com}.}
    \label{fig:light_curve}
\end{figure}

The simulated exopet, represented by a chosen template, is initially outside the confines of the rectangle, centered at some $x_0$. It is clear that $f(0)=1$. We divide the transit into discrete timesteps of duration $\Delta t$. After each timestep, the exopet's position is changed by some $\Delta x$, meaning that after $k$ timesteps, its center lies at $x_k=x_0+k v\Delta t$, with $v$ the speed of the exopet, assumed to be constant. The flux is then calculated, and the procedure is repeated until the transit has finished. Figure \ref{fig:light_curve} shows a sample projection template for a cat — an example of an exopet — and the light curve formed by its subsequent transit in front of a webcam.

This simple simulation is performed under the assumption that exopets behave similarly to exoplanets - that is, during each transit, they enter and exit the field-of-view only once, and remain in sight for predictable amounts of time. However, this is oversimplistic and fails to accurately explain prior exopet transits. We instead propose a more complex seven-phase model:

\begin{enumerate}
    \item[1.] \textbf{Ingress}, as the exopet enters view.
    \item[2.] The \textbf{main portion of the primary transit}, as the exopet remains in front of the Zoom screen.
    \item[3.] \textbf{Egress}, either by the pet's own volition or via removal by its host astronomer.
    \item[4.] \textbf{An apology} for the interruption by the host astronomer, followed by responses from other Zoom attendees pointing out that exopets are absolutely adorable and asking for an encore.
    \item[5.] The \textbf{return} of the exopet.
    \item[6.] The \textbf{main portion of the secondary transit}, which may last anywhere from seconds to tens of minutes, depending on how easily distracted the exopet is and the length of the Zoom call.
    \item[7.] \textbf{A second egress}.
\end{enumerate}

\begin{figure}
\centering
	\includegraphics[width=0.8\columnwidth]{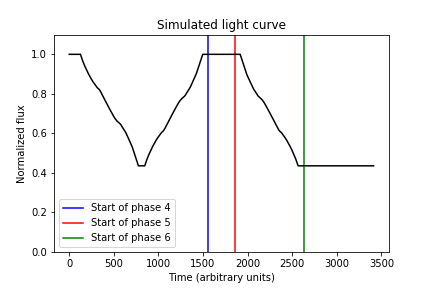}
    \caption{A more realistic exopet transit event has a secondary transit, consisting of several additional phases. The main difference is that the point of maximum transit likely lasts longer, as the exopet settles in front of the camera and gets petted. The duration of this secondary transit may be long, and so we omit the final egress from this diagram.}
    \label{fig:extended_transit}
\end{figure}

A simulation of this seven-phase model is presented in Figure \ref{fig:extended_transit}. We note that it may be difficult to determine actual light curves of exopets due to the challenge of obtaining time series photometry (i.e., recordings of Zoom meetings).

\section{Data Collection}\label{sec:obs}

%briley can describe google form design here and put cleaned data table in appendix

To obtain as much data as possible on the new phenomenon of exopets, we gathered observations from multiple observers via the crowd-sourcing using Twitter\footnote{\url{https://twitter.com/}} and Google Forms\footnote{\url{https://www.google.com/forms/about/}}. The survey respondents provided information on the color, floofiness (on a scale of 1-5), type of animal, percentage of Zoom screen covered, and duration of the event. We encountered a greater variety of exopet events than originally expected, including not only dogs and cats but also a pig, a goat, and three birds. Observers also captured not only transit events, wherein the pet walks in front of the camera, but also microlensing events where the camera is directly pointed at the pet. Transit observations are discussed in Section \ref{sec:transits} and microlensing is discussed in Section \ref{sec:microlensing}. Some observers were also able to provide follow-up direct imaging data, discussed in Section \ref{sec:imaging}. We follow typical naming conventions for new exop(lan)etary objects, e.g. ``HostStar b'', ``HostStar c'', and so on, where the host star is the human associated with the pet. Some of these objects have a colloquial name as well, which will be referred to in parentheses after the conventional name.

A summary of our data is visualized in Figure \ref{fig:Normal_transits}. We see a Gaussian-like distribution of floofiness, centered around a floofiness score of 3. Transit depth is skewed towards smaller values, and transit duration is similarly skewed towards short transits. We have not analyzed the completeness of the data in order to determine if these are genuine pup-ulation statistics, or artifacts of our observational method; this analysis is outside the scope of this paper. However, there is still a great deal of information that we can glean about exopets given the large amount of data we have collected from multiple observation methods. See the Appendix for the full data set and Figure \ref{fig:transitpics} for examples of exopet transits on Zoom. The following sections focus on analysis and interpretation of the data.

\begin{figure}
    \centering
    \includegraphics[width=\linewidth]{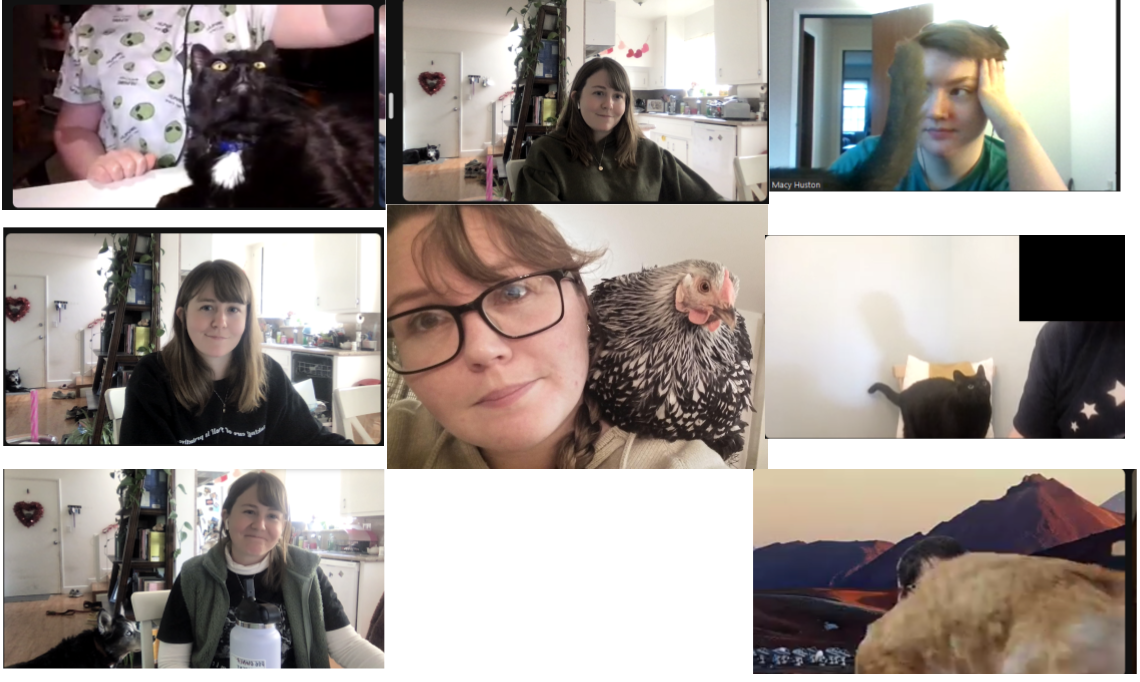}
    \caption{Examples of exopet transits on Zoom.}
    \label{fig:transitpics}
\end{figure}

\section{Observations and Results}\label{sec:results}

Here we discuss our results from transits, microlensing, and imaging observations. Each of these observation methods provides different insight into the phenomenon of exopets on Zoom, and their various characteristics: mass, radius, orbit, color, floofiness, etc.

\subsection{Transits}\label{sec:transits}
%Sabina and Ali

An exopet whose appearance causes it to transit the face can be detected by monitoring the stellar flux for dips. At the most basic level, transit events can be characterized in terms of their depth, duration, and probability of being observed. The transit technique has demonstrated a sensitivity to exopets ranging from super-Fluffs to mini-Pets appearing in front of a diverse array of faces. 

We obtain the exopet length from Equation \ref{pet_radius}. We take $R_{\smiley{}}$ as the average height of a sitting h\textit{oo}man, which is, according to \cite{Potkany2018}, 86.25 $cm\approx 1.35\times 10^{-7} R_E$ (we take the average across the species). Here $R_E$ is the radius of the Earth.
%%% we should make sure the math here aligns with the math in the transit theory section --Briley

We normalize the length of an exopet by $R_{\smiley{}}$, taking the value for the face radius as 1. 

The parameter set for every exopet $n$ is presented as $\textbf{p}_n=$[floofiness (F), depth ($\delta$), duration (T), length ($L_{paw}$)], as defined in Section \ref{sec:theory}. We normalize floofiness by Boston Terriers, whose value of floofiness is 1. Floofiness therefore ranges from $0$ to $5$, where 0 is a Sphynx and $5$ is a Samoyed. 

There are $N$ exopets, each of which has $j$ parameters ($1\leqslant j\leqslant 4$), which give the floofiness spectrum measurements $F_{nj}$ for every exopet $n$ ($1\leqslant n\leqslant N$). We calculate the total weighted standard deviation for each parameter $j$, which is given by

\begin{equation}
    \sigma_j=\left(\sum_{i=0}^N \sigma_i \right)^{-1/2}
\end{equation}\label{eqn:std}

\begin{figure}
\centering
	\includegraphics[width=0.4\columnwidth]{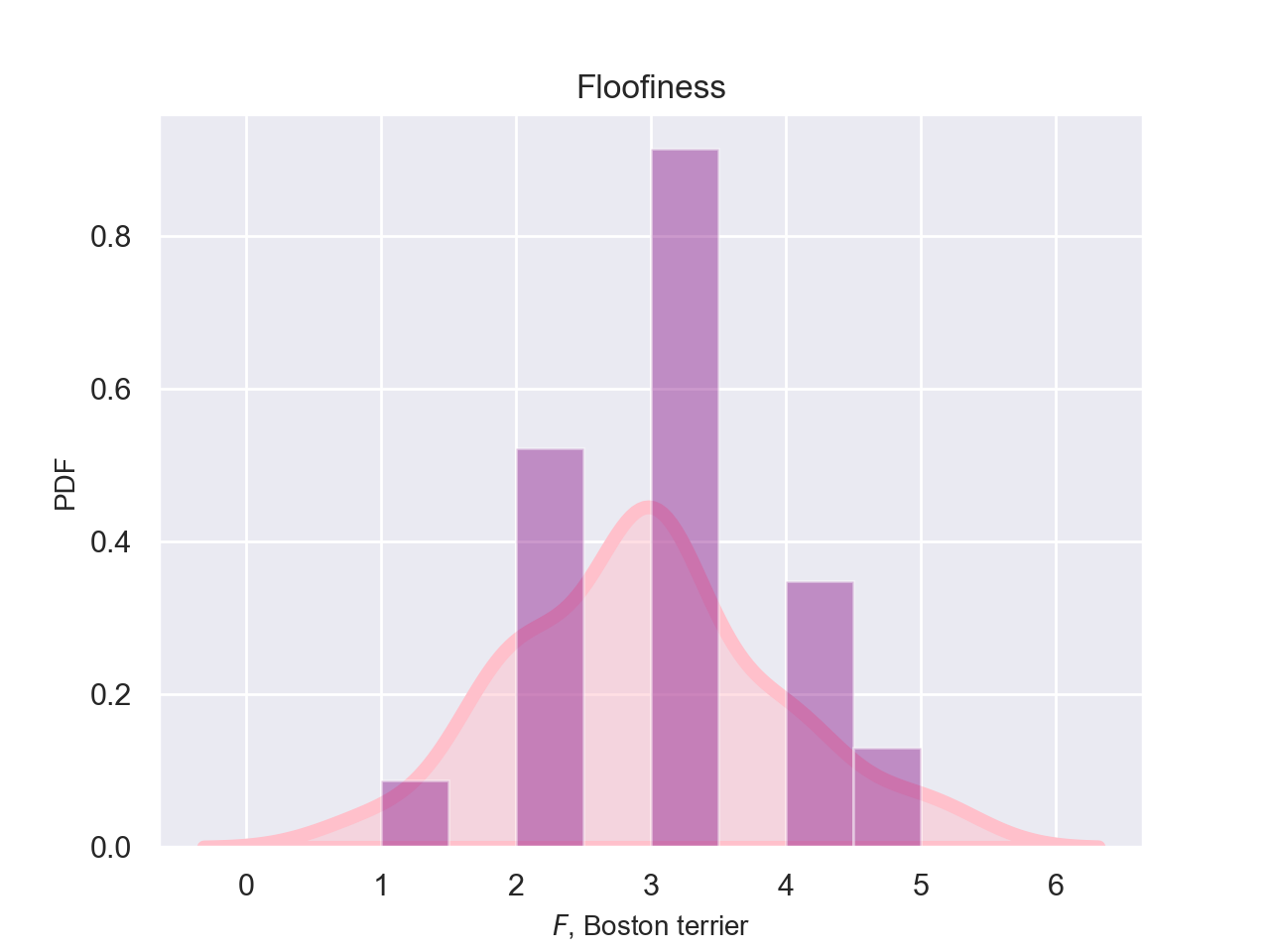}
	\includegraphics[width=0.4\columnwidth]{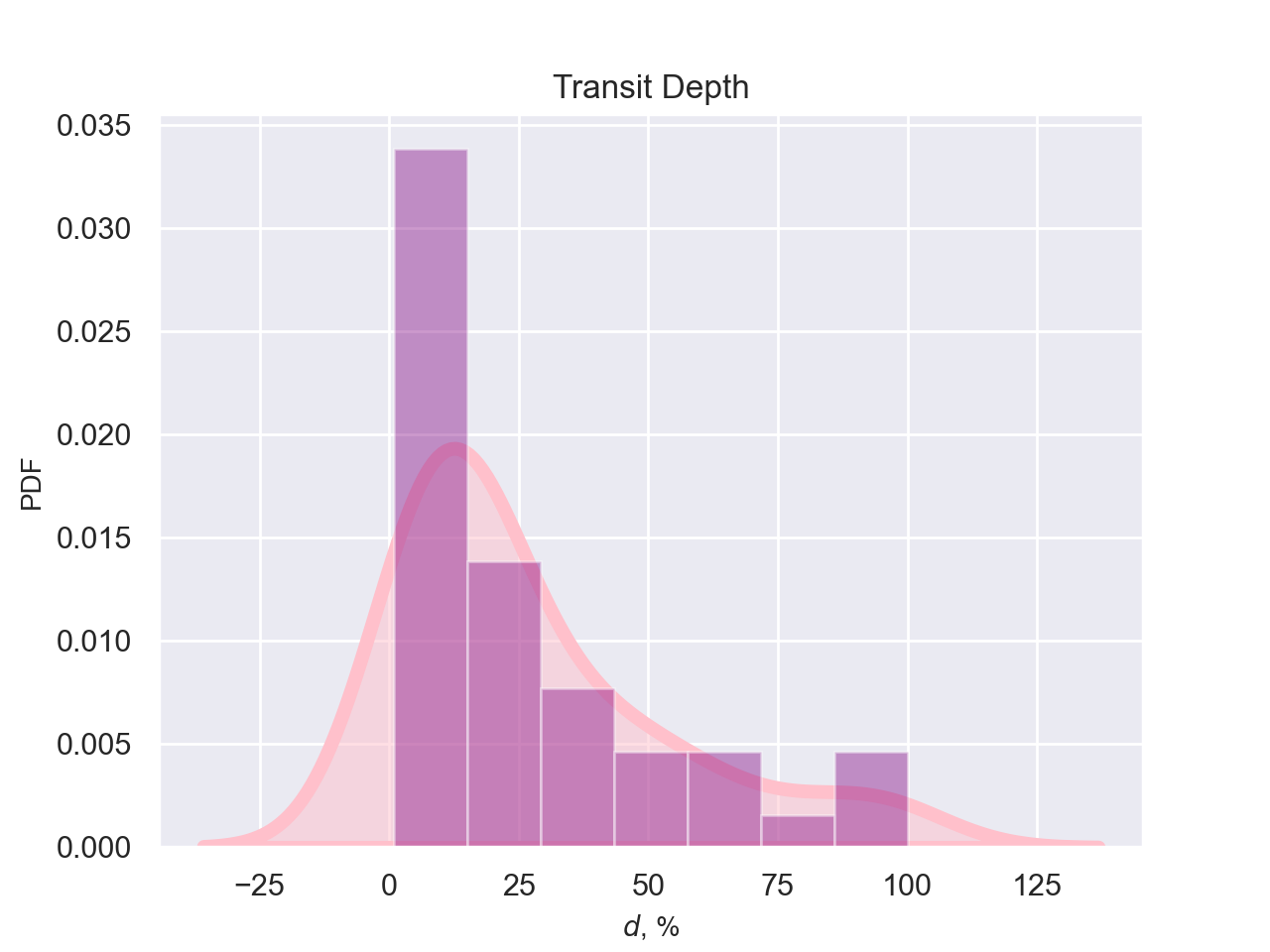}
	\includegraphics[width=0.4\columnwidth]{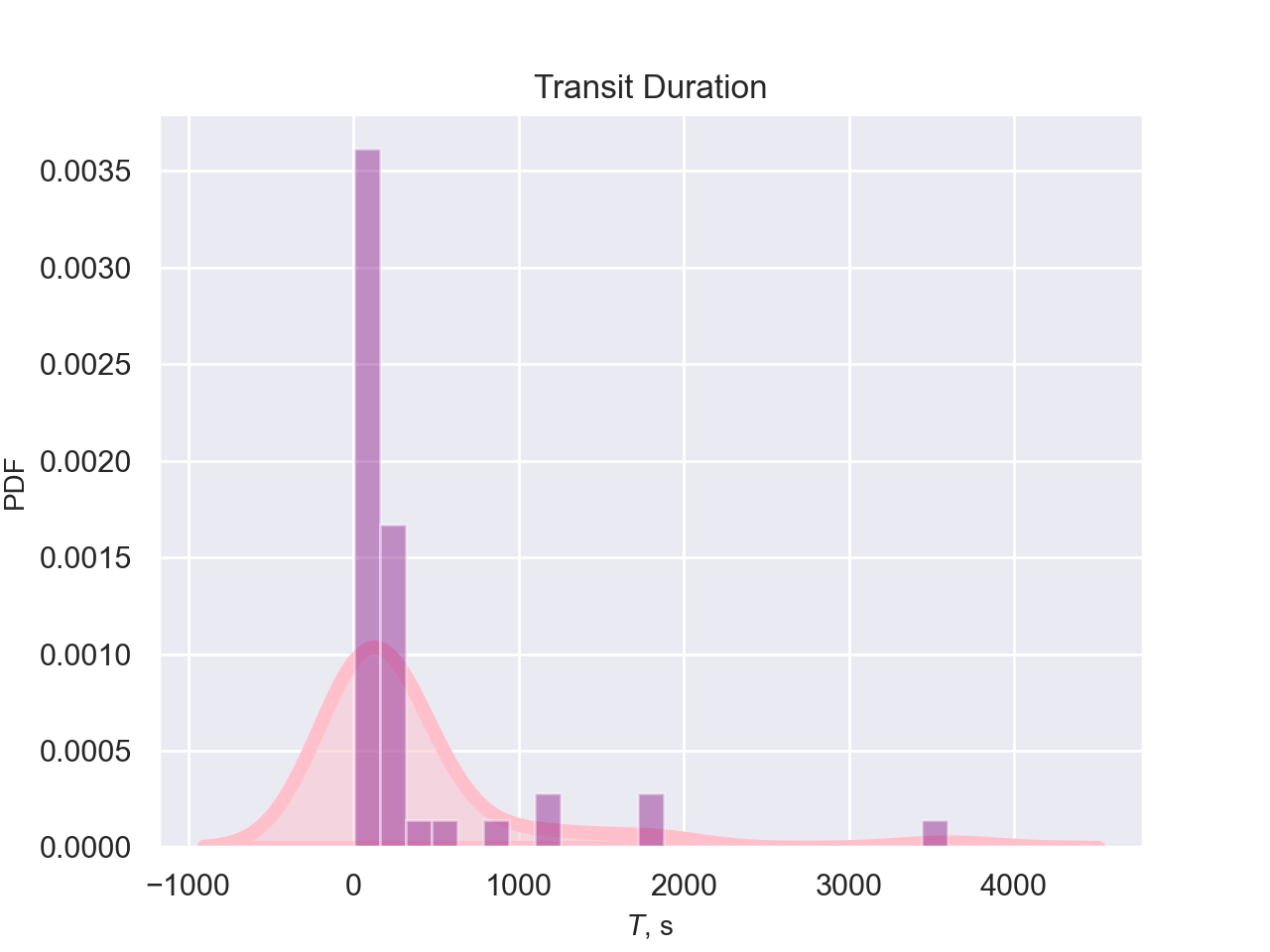}
	\includegraphics[width=0.4\columnwidth]{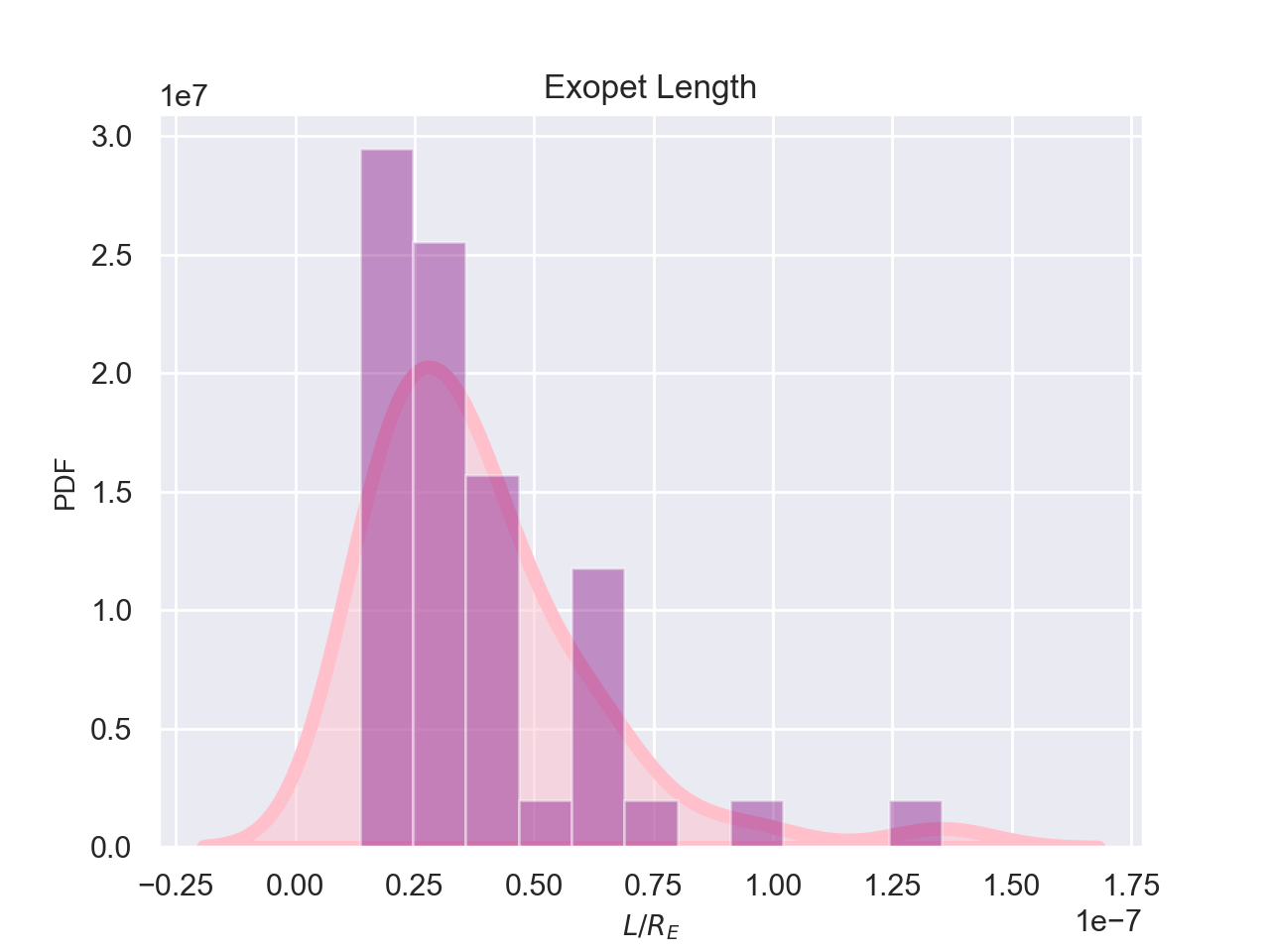}
    \caption{Probability Distribution Functions (PDFs) of the parameter-space $\textbf{p}_n$ for the $j^{th}$ parameter of an $n^{th}$ exopet.}
    \label{fig:Normal_transits}
\end{figure}

We fit our parameter data with a normal distribution, the results of which are shown in Figure \ref{fig:Normal_transits}. The standard deviations are $\sigma_{nj}=\{0.93,26.11,644.90,2.32\times 10^{-8}\}$ for the $j^{th}$ parameter in parameter-space $\textbf{p}_n$. An increase in weighted mean can be interpreted as excess absorption in the spectral feature as suggested by \cite{Feinstein_2021}.

\begin{figure}
\centering
	\includegraphics[width=0.4\columnwidth]{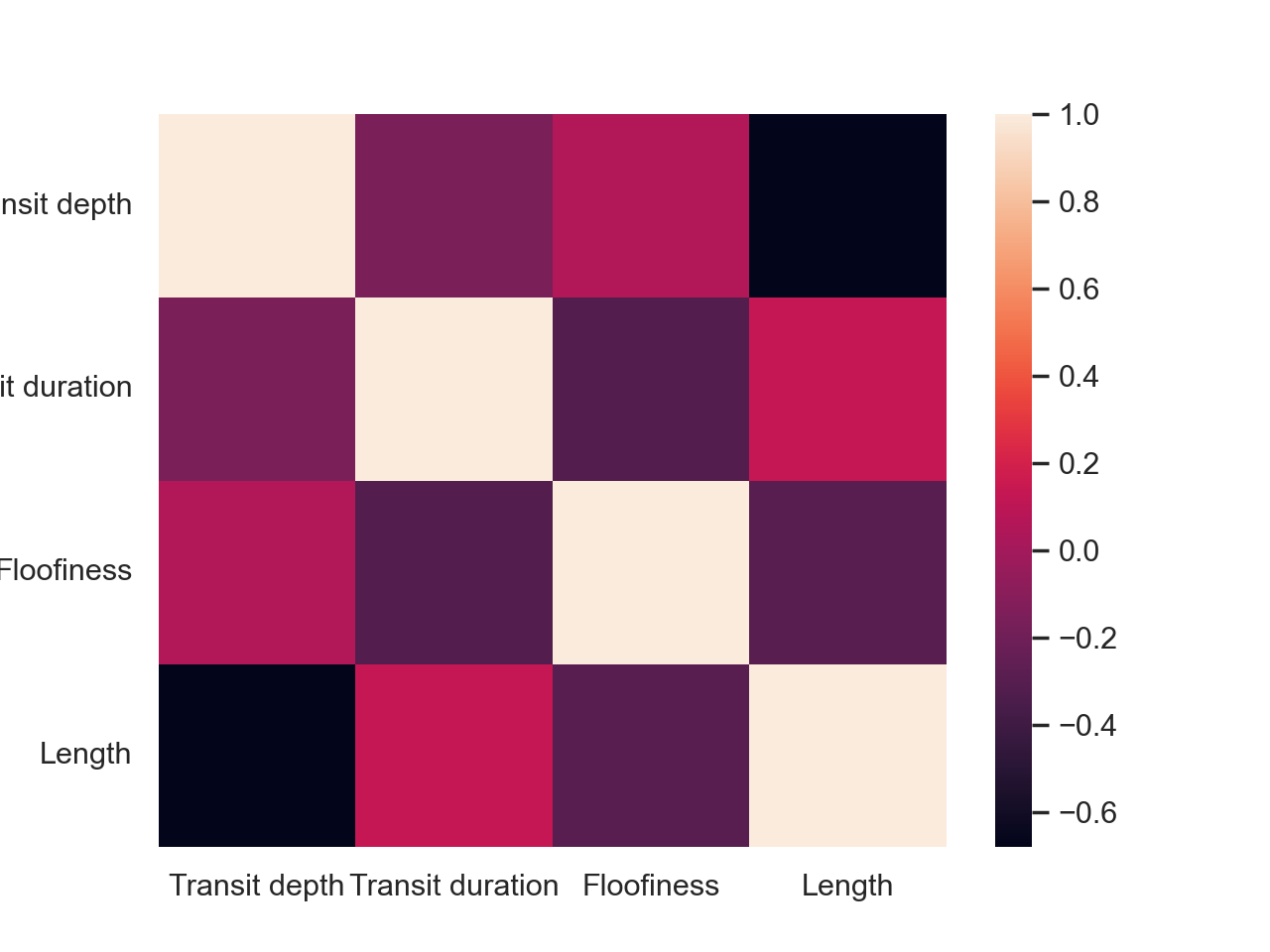}
    \caption{Spearman’s correlation value for floofiness, transit depth and transit duration. The average Spearman’s value across all spectral features is -0.18, indicating that there is no correlation.}
    \label{fig:heatmap}
\end{figure}

The more floofy an exopet is, the smaller $R_{\smiley{}}$ is; thus the flux that the Zoom camera receives gets smaller following Equation \ref{pet_radius}. We therefore test for correlation between the parameters for floofiness $F_{nj}$, for transit duration $T_{nj}$, and for transit depth $\delta_{nj}$ in Figure \ref{fig:heatmap} by calculating the Spearman’s rank correlation coefficient \citep{Spearman1907}. This test assesses how well two data sets can be described using a monotonic function. A correlation coefficient of $\pm 1$ implies an exact monotonic relationship, while a coefficient of 0 implies no correlation.

\begin{figure}
\centering
	\includegraphics[width=0.8\columnwidth]{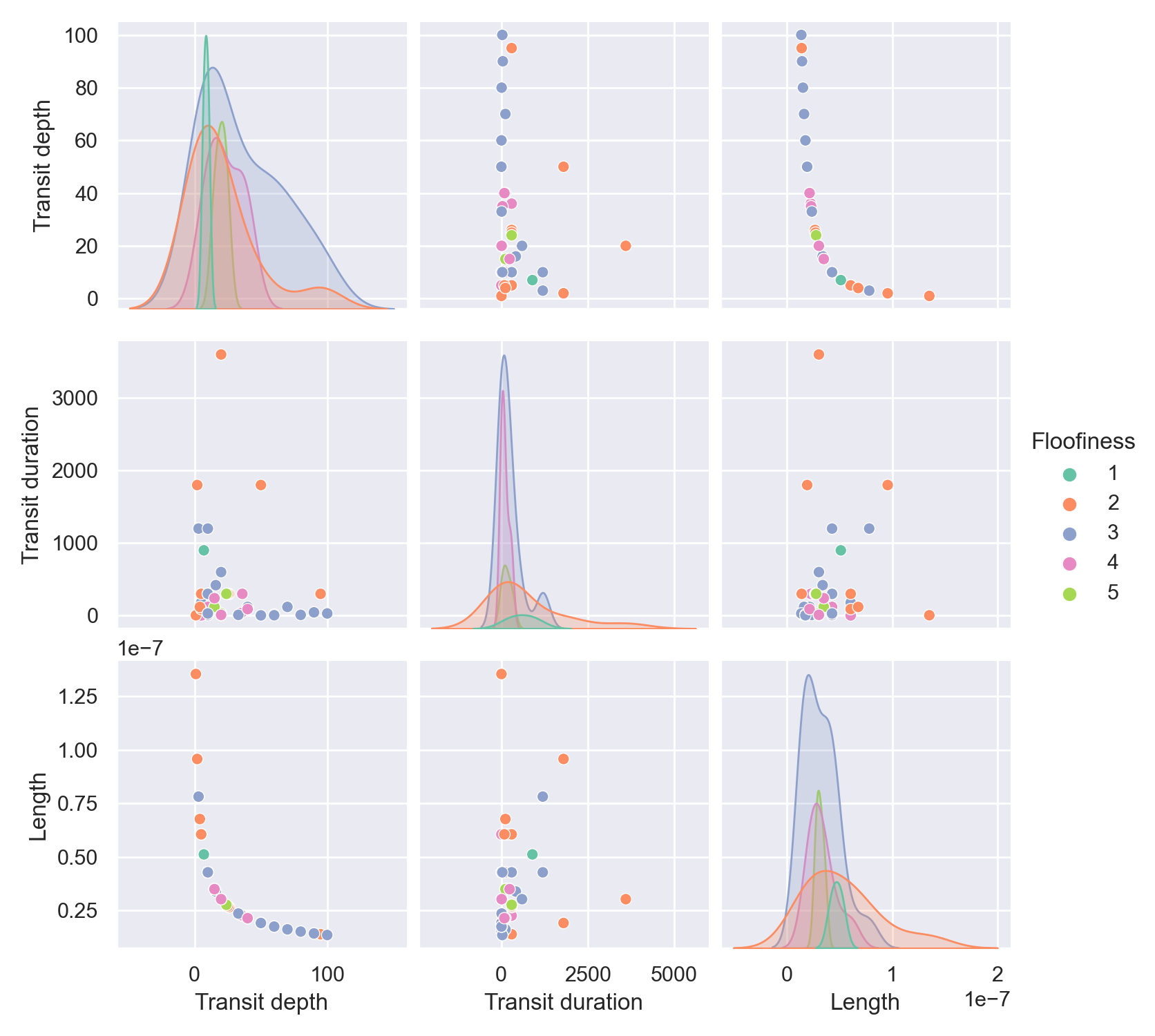}
    \caption{The correlation between parameters $T_{nj}$ and $\delta_{nj}$ with floofiness $F_{nj}$. The plot shows that the largest transit depth corresponds to $F_{nj}=1$ and $F_{nj}=3$, which contradicts the initial hypothesis. For transit duration, the longest $T_{nj}$ corresponds $F_{nj}=3$ and the shortest to the value $F_{nj}=5$.}
    \label{fig:pairplot}
\end{figure}

We test how floofiness affects transit depth, since we hypothesize that more floofs should cover more faces. We find correlation values of 0.06, -0.32, and -0.29 for transit depth, transit duration, and exopet length respectively, which implies that there is almost no correlation between transit depth and floofiness. Transit duration and floofiness, and exopet length and floofiness, are slightly anti-correlated. We also show in Figure \ref{fig:pairplot} that the largest transit depth corresponds to a floofiness value of 1, which contradicts our hypothesis. For the transit duration, however, we find that the longest transit corresponds to a floofiness value of 3, which also corresponds to the longest exopet length. This may be due to \textit{snuggliness} of an exopet. Exopets of floofiness 3 may have a tendency to snuggle with the hooman, which therefore increases their flux in Zoom cameras; however, this analysis is beyond the scope of this paper.


\subsection{Microlensing Events}\label{sec:microlensing}

% MH - oof microlensing is hard to make a great analogy with but i will do my best. i'm going to say transit depth = planet/star mass ratio, since that's a measureable quantity from microlensing (is it worth implementing a mass-radius relationship here?- probably not, i have a lot to do this week). i'll say duration is einstein ring crossing time. can't think of a microlensing analogy for floofiness, so let's say that comes from direct imaging follow up. 

Microlensing events occur when a brief serendipitous physical alignment, typically of a planet or passing body, serves to magnify a source in the observer's view \citep{griest1998use}. This has served as a planet detection technique which is sensitive to exoplanets more distant from their hosts than those typically detected via the transit and radial velocity methods, as well as lower-mass exoplanets \citep{tsapras2018microlensing,Gaudi2012}. %the lower mass point may not be interesting, since we have observed birds in both transit and microlensing, so I can't really make a point about it with our data.
In the field of exopets, microlensing events are defined differently. They are similarly brief, but instead of a serendipitous alignment of bodies, it is a deliberate alignment of the field-of-view of the camera to the exopet in question. While some may question the validity of an exopets detection method that is predicated by the deliberate action of it's host, we point out that inherent in our methodology was never to request that the exopet be shown on screen, and so in that way the microlensing event was serendipitous. The motivation and willingness of an exopet host to show off their exopet is outside the scope of this research. % MH - I was considering just kind of glossing over caveats about the imperfect analogy here, but this works too if y'all prefer it CC - I love it!

\begin{figure}
    \centering
    \includegraphics[width=.49\textwidth]{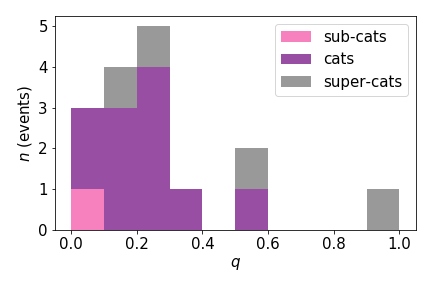}
    \includegraphics[width=.49\textwidth]{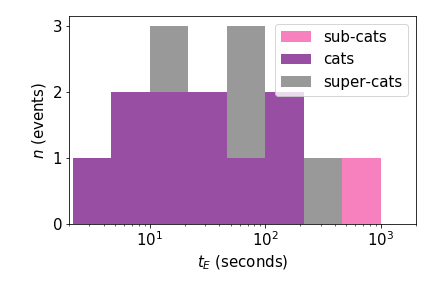}
    \caption{Microlensing event statistics. \textbf{Left}: Exopet-to-host mass ratios. \textbf{Right}: Timescales. Colors indicate exopet type.}
    % MH - I changed the color scheme to try to coordinate better with the other figures. This should be color-blind friendly, according to someone on github
    \label{fig:micros}
\end{figure}

Even though they are defined differently, we still attempt to use some of the theoretical framework from exoplanet microlensing \citep{Gaudi2012} in the emergent field of exopets. When a host aligns their camera with an exopet, we observe its gravitational influence as some fraction of its host's
\begin{equation}
q = m_{\rm pet}/m_{\rm host},
\end{equation}
over some timescale
\begin{equation}
t_E = \mu_{\rm rel}/\theta_E, 
\end{equation}
where $m_{\rm pet}$ is the exopet's mass, $m_{\rm host}$ is the host's mass, $\mu_{\rm rel}$ is the exopet's proper motion across the screen, and $\theta_E$ is its angular Einstein ring radius. We observed 16 microlensing events and acquired direct imaging follow-up for 6 of them. The distributions of mass ratio and timescale are shown in Figure \ref{fig:micros}.

While we recognize that this is a relatively small sample size, we can comment on some of the larger trends. We conclude from the mass ratio ($q$) distribution that most hosts tend to have pets that are approximately one third their size or smaller. This is regardless of the currently proposed sub-cat, cat, and super-cat categories. This can be interpreted in a few ways. The first is that the current system for categorizing exopets does not sufficiently describe the true diversity of the phenomena, leading to observer bias. By focusing the field's language on a single popular well known pet in an attempt to simplify or popularize the field, we are instead necessarily washing away the dramatic range of exopet possibilities. Secondly, exopet hosts are perhaps more diverse than first appreciated. In Section~\ref{sec:transits} we assumed that all exopet host faces were approximately the same size. The microlensing results indicate that there is a larger diversity in exopet hosts than previously thought. These results also indicate that the ratio between the mass of an exopet and its host is similar for all categories of exopets, since hosts of a given size favor exopets of a given size. There is a similar analogy in exoplanets. While diverse types of exoplanets have been found over a large range of stellar hosts, certain types of exoplanets are more likely to be found orbiting certain types of stars 
\citep{trappist,adibekyan_2021}. This interpretation can also be useful in explaining the outlier point where a super-cat exopet and its host have a mass ratio of 1. While additional follow-up is needed, it may be possible to consider this super-cat human-sized, or the human super-cat-sized.

\subsection{Direct Imaging Follow-Up}\label{sec:imaging}

Although the transit and radial velocity methods have been more prolific in the discovery of new exoplanets, the technique of direct imaging provides insight into a population of planets that indirect methods cannot yet reach --- those on wider orbits that are farther from their host stars \citep{marois2008direct}. Additionally, direct imaging provides us with mass, a fundamental parameter of an exoplanet, and useful information about an exoplanet's composition and orbit \citep{oppenheimer2009high, traub2010direct}. In exoplanet science, this technique has generated incredible discoveries, revealing not only fully-formed planets, but also their earlier formation stages. Some examples of these discoveries are the HR 8799 multi-planet system, the $\beta$ Pictoris system with both an imaged exoplanet and a debris disk, the forming planet PDS 70 b still embedded in its disk, and the lowest-mass exoplanet imaged to-date: 51 Eridani b \citep{marois2008direct,absil2013searching,keppler2018discovery, macintosh2015discovery}. 

Direct imaging of exoplanets is somewhat analogous to direct imaging of exopets, but there are a few key differences. An exoplanet image must be acquired with significant hardware and software infrastructure (e.g., by some estimates, the Magellan Telescopes in Chile operate at expenditure of $1$ s$^{-1}$), all to solve the problem of blocking out light from a bright host star \citep{macintosh2014first}. An exopet image, on the other hand, is much easier and less expensive to obtain since the targets are much nearer to the observers, creating a greater angular separation from the host human. Additionally, humans are approximately as luminous as exopets, so there is not a huge contrast differential required to image an exopet like there is for an exoplanet.

\begin{figure*}\label{fig:images}
    \centering
    \includegraphics[width=0.6\linewidth]{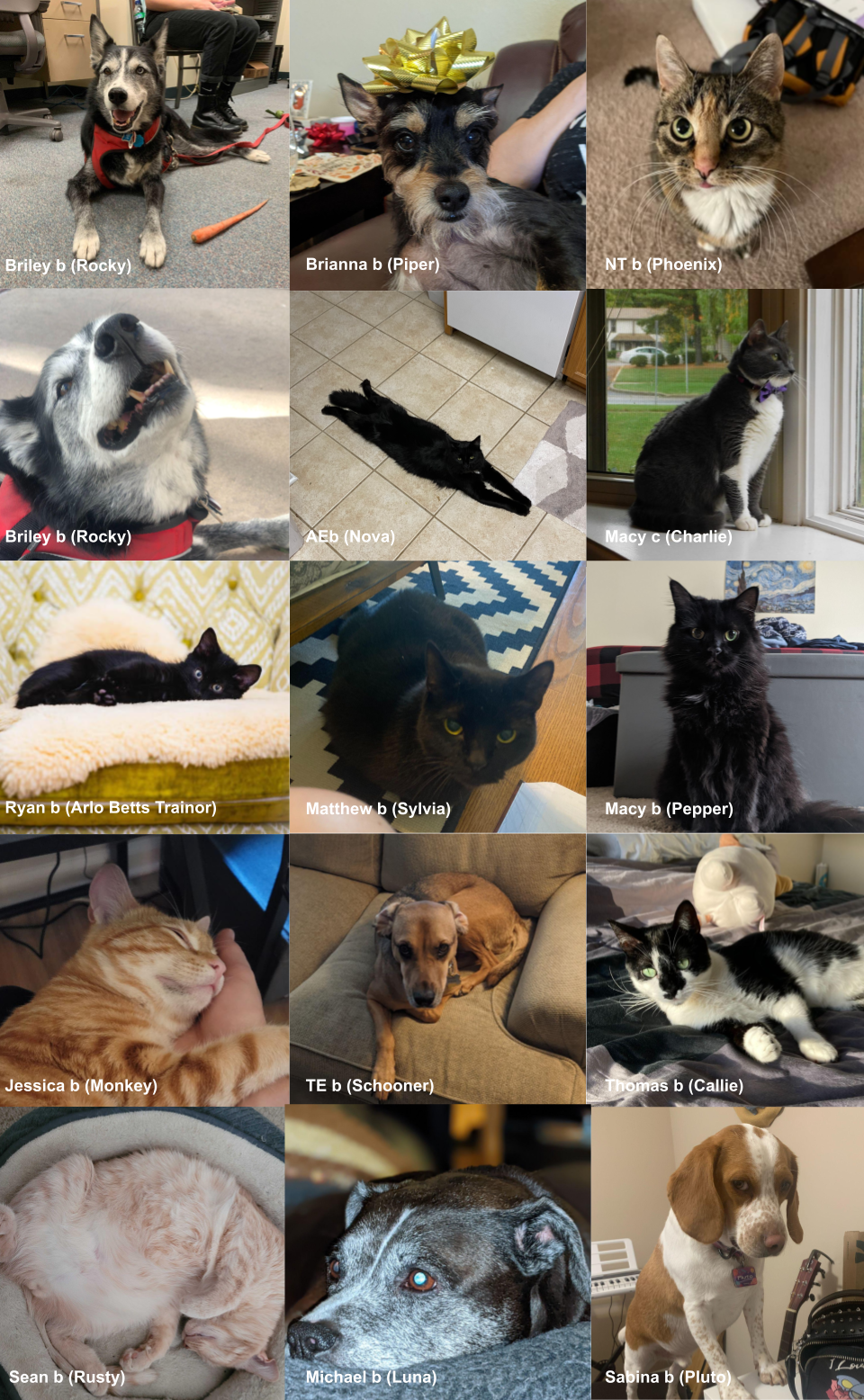}
    \caption{Direct imaging data for the exopets in our sample (wherever applicable).}
\end{figure*}

We are able to image exopets of all masses, as illustrated in Table \ref{tab:SNR}. This table also includes data on the signal-to-noise ratio (SNR) measured in each exopet image. In general, we notice that the SNR when imaging exopets is dependent on the color of the pet and the color of their environment. Whereas a complication when imaging exoplanets is the high contrast of a small planet orbiting a bright star, the problem when imaging exopets is \textit{needing} a high contrast between the pet and its environment. All exopet direct imaging data is shown in Figure \ref{fig:images}.

Exoplanet imaging relies on formation models to derive a mass based on the observed brightness of the exoplanet \citep{chabrier2000evolutionary,baraffe2003evolutionary,spiegel2012spectral}. Similarly, exopet imaging relies on formation models to determine mass. Exoplanets cool with time, as they radiate away the heat from their formation, but exopets grow with time as they age. Their size function therefore increases more rapidly at younger ages, and then plateaus at older ages in the exopet's life. By measuring the size of the exopet, we can both estimate its mass and age, although age measurements are often unreliable at older ages. Additionally, unlike exoplanets, exopets often have different ages than their host humans, adding an additional layer of complexity to this problem. It is worth noting that the correct formation model must be selected depending on the type of exopet, since cats, dogs, birds, etc. have very different formation tracks. Mass estimates are also listed in Table \ref{tab:SNR}.

\begin{figure}[h]
    \centering
    \includegraphics[width=0.48\linewidth]{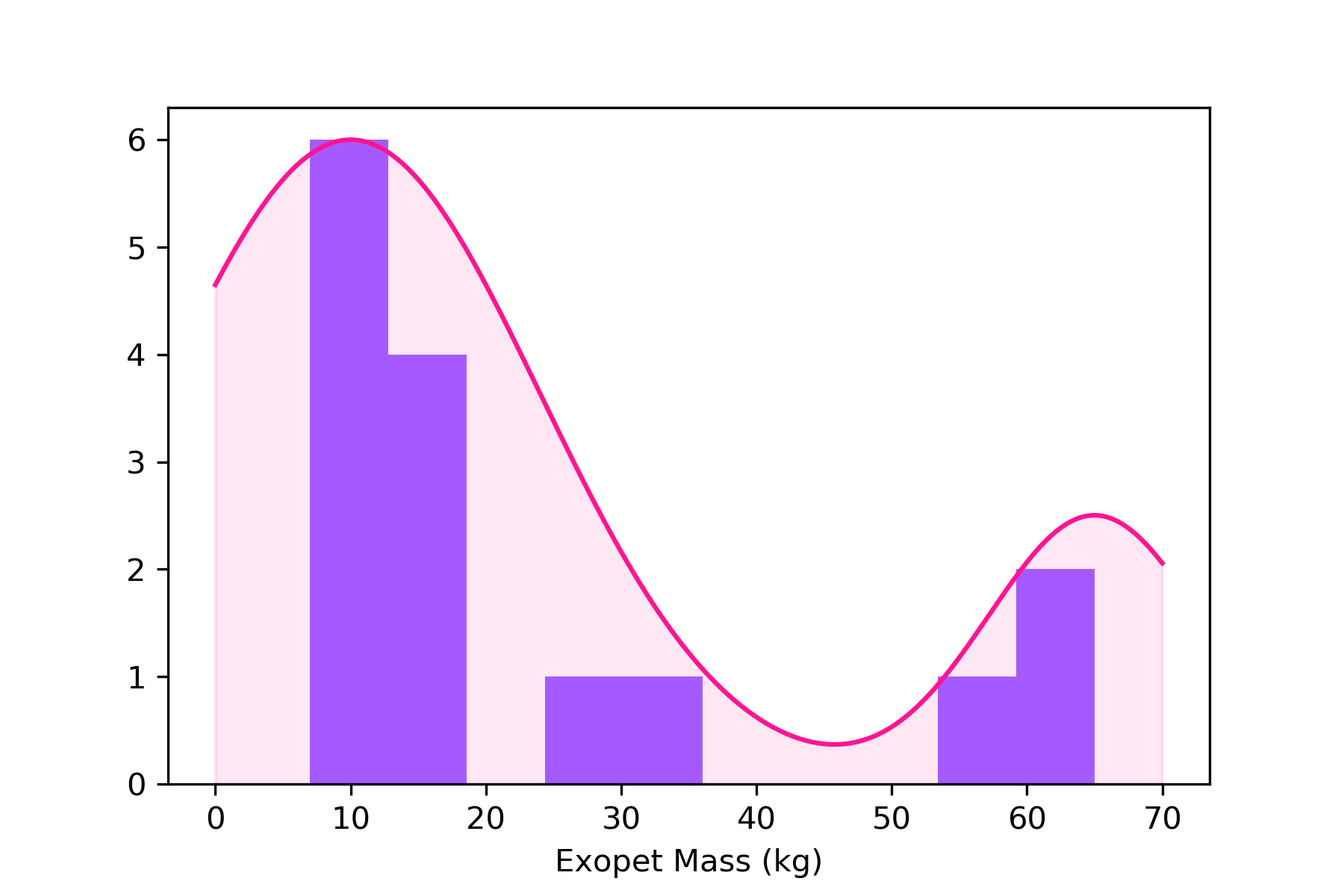}
    \includegraphics[width=0.48\linewidth]{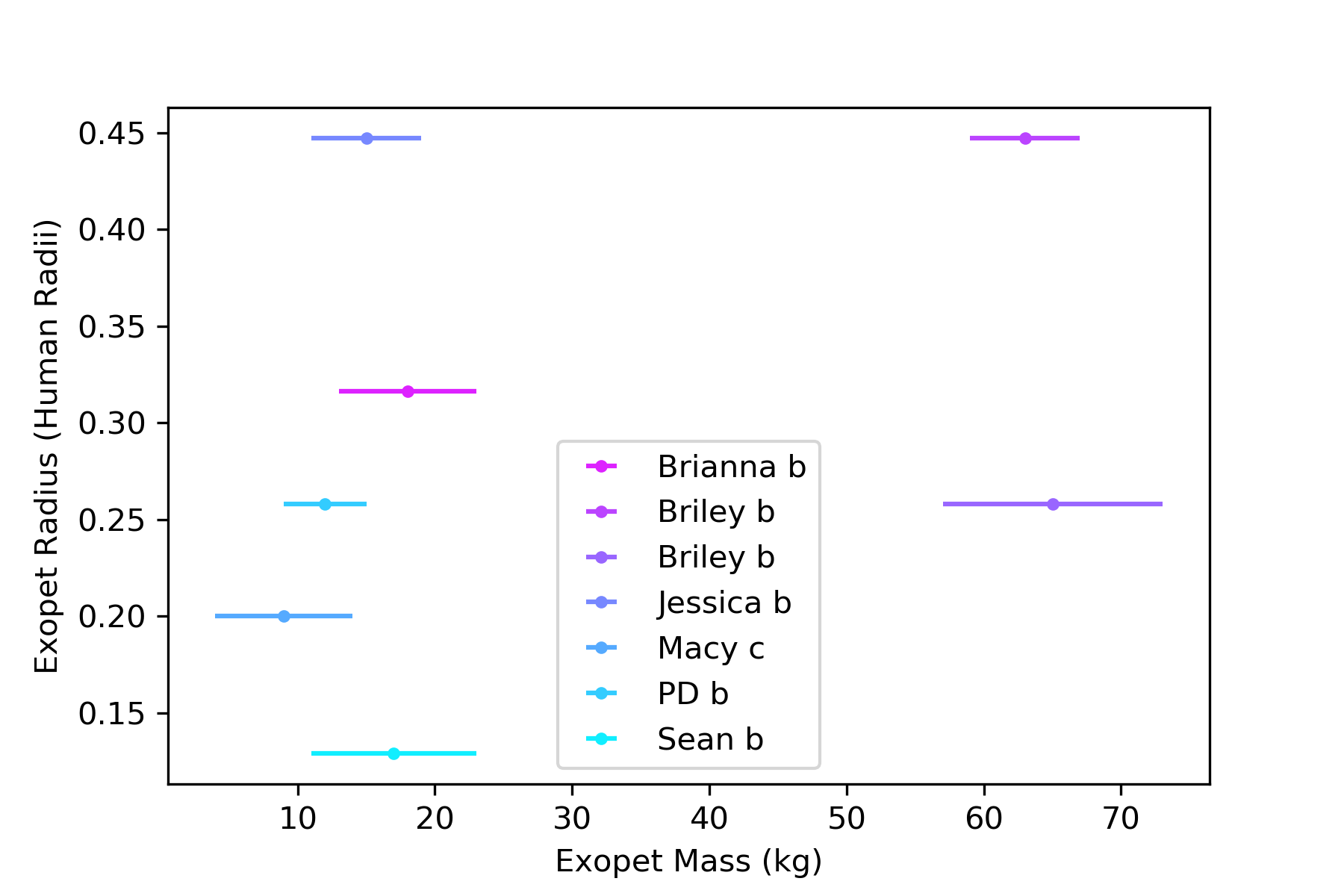}
    \caption{\textbf{Left:} Mass distribution for exopets, showing a bimodal distribution with peaks around $\sim10$ and $\sim60$ kg. Peaks correspond to two distinct populations -- cats and dogs, respectively. \textbf{Right:} Exopet mass-radius diagram, showing that, unlike exoplanets, there is little correlation between mass and radius. It is worth noting that our sample is small and the mass errors are relatively large, so future work may reveal correlations to which we are not yet sensitive.}\label{fig:direct-masses}
\end{figure}

In Figure \ref{fig:direct-masses}, we explore the mass distribution of observed exopets. Our best fit to the data is a multi-modal distribution with two peaks: one around 10kg, which we believe is the exocat population, and one around 60kg, which we believe is the exodog population. We also combine these mass estimates with radii from our transit data (Section \ref{sec:transits}) to produce a mass-radius plot. Although our sample of exopets that have both measured masses and measured radii is small, from this initial investigation we see no correlation between exopet radius and exopet mass. This is in contrast to exoplanet observations, which have shown groupings based on exoplanet composition \citep{swift2011mass}.

% \usepackage{longtable}
% Note: It may be necessary to compile the document several times to get a multi-page table to line up properly
% \usepackage{longtable}
% Note: It may be necessary to compile the document several times to get a multi-page table to line up properly
\begin{longtable*}[c]{|c|c|c|c|}
\hline
\textbf{Target} & \textbf{Observation Date} & \textbf{SNR} & \textbf{Mass (1.6 $\times$ 10$^{-25}M_{\bigoplus}$} \\
\hline
AE b (Nova)                 & Feb-14 & 10.2 & 10$\pm$3 \\ \hline
Brianna b (Piper)           & Feb-16 & 4.8 & 18$\pm$5\\ \hline
Briley b (Rocky)            & Feb-23 & 6.1 & 63$\pm$4 \\ \hline
Briley b (Rocky)            & Feb-23 & 8.2 & 65$\pm$8 \\ \hline
Jessica b (Monkey)          & Feb-8  & 5.3 & 15$\pm$4 \\ \hline
Macy b (Pepper)             & Mar-8  & 7.8 & 8$\pm$6 \\ \hline
Macy c (Charlie)            & Feb-9  & 5.9 & 9$\pm$5 \\ \hline
Matthew b (Sylvia)          & Feb-9  & 4.9 & 26$\pm$8 \\ \hline
Michael b (Luna)          & Feb-9  & 5.5 & 11$\pm$3 \\ \hline
Ryan b (Arlo Betts Trainor) & Feb-14 & 8.4 & 7$\pm$2 \\ \hline
NT b (Phoenix)              & Mar-8  & 6.0 & 14$\pm$4 \\ \hline
Sabina b (Pluto)            & Mar-23  & 7.1 & 31$\pm$7 \\ \hline
Sean b (Rusty)              & Feb-9  & 5.2 & 17$\pm$6 \\ \hline
TE b (Schooner)             & Feb-16 & 5.4 & 55$\pm$10 \\ \hline
Thomas b (Callie)           & Feb-8  & 5.7 & 8$\pm$4 \\ \hline
\caption{Signal-to-noise ratios (SNR) and mass estimates for directly imaged exopets.}
\label{tab:SNR}\\
\end{longtable*}

\section{Discussion} \label{sec:disc}

To our knowledge, this work represents the first thorough study of its kind to understand and characterize the exopet phenomenon. However, we recognize that our analysis is not complete and there remains scope of a number of follow-up observations and analyses. In this section we discuss two kinds of exotic exopets that we know our methodology was insensitive to, as well as attempt an initial estimate of the number of total exopets based on our work.

\subsection{Exotic Exopets}\label{sec:exotic}

While this work is mostly focused on the more commonly transiting exopets, we do point out other known exopets that are more difficult to observe via the methods discussed above. 

\subsubsection{\textbf{S}ub-\textbf{N}eptunian \textbf{A}nimal \textbf{K}eplerian \textbf{E}xtended Bodies}

In Section \ref{sec:transits} we make assumptions on the general size and shape of exopets. However, the class of exopets known as Sub-Neptunian Animal Keplerian Extended bodies (SNAKEs) break some of these basic assumptions. The body of SNAKEs are often elongated beyond the regular proportions we previously estimated, such that one axis may be an order of magnitude longer than the others. Two examples may be seen in Figure \ref{fig:snakes}. This exopet shape may be caused by one or a series of tidal interactions during the formation process, but this theoretical work is beyond the scope of this paper.

This elongation could cause significant deviations from the the transit model used in previous sections. While we do not extensively model the light curve, we note that the orientation of the SNAKEs will cause a dramatic change in the possible transit depth, depending on if the extended axis is along the observers line of sight or not. Although we should start from an assumption of an istropic distribution of orientations, if there is some preferred orientation for the extended axis to align with the Zoom line of sight, than the transit depth would be far more difficult to detect. We suggest the field commit to a thorough, high resolution, long baseline survey to better constrain these potential transiting events. There may well be thousands of SNAKEs that go uncounted in every Zoom call.

Furthermore, observations of SNAKEs are often polarized when compared to other exopets. The polarization occurs such that they are either warmly received or dramatically rejected by the observer. This would likely negatively influence the number of microlensing events (see Section~\ref{sec:microlensing}) initiated by hosts. This observer bias must be considered another factor as to why no SNAKEs were reported in our sample, and we suggest future should be focused on better understanding this effect.

\subsubsection{\textbf{D}ynamically \textbf{U}nstable \textbf{C}oplanar \textbf{K}epler Objects}

Another notable outlier which has been difficult to reproduce with our model predictions outlined in Section \ref{sec:theory} are Dynamically Unstable Coplanar Kepler objects (DUCKs). These DUCKs feature highly unstable and unpredictable orbits that can be nearly vertical in the Zoom reference frame. Qualitative descriptions of in-situ observations of individual and small groups of DUCKs demonstrate this instability, with descriptors including ``taking flight'' and ``flying across the screen''. In addition to impacting the exopet orbit, this also affects the frequency of transit, making DUCK orbits challenging to predict on both short and long timescales.

Still, DUCKs do share some similarities with the well-characterized larger sample of exopets. For example, our detailed in-situ observations of one small group of DUCKs (see Figure \ref{fig:in_situ_DUCKs}) demonstrate their color is Brown, like many of the more common exopets in our sample. However, their Floofiness index is hard to quantify robustly using the standard methods adopted in this paper. Follow up multiwavelength and/or spectroscopic observations are required to constrain the global properties of DUCKs and further develop our exopet model to include DUCKs and other exotic transiting objects.

\subsection{The \textbf{D}omesticly \textbf{R}esiding \textbf{A}nimal \& \textbf{K}itty \textbf{E}stimation (D.R.A.K.E.) Equation} %we can just make an equation based off the drake equation but maybe not try to come up with an answer to it.

Following the framework of the Drake Equation \citep{drake1965}, we propose the \textbf{D}omesticly \textbf{R}esiding \textbf{A}nimal \& \textbf{K}itty \textbf{E}stimation. While the Drake Equation estimates the number of actively communicative extraterrestrial civilizations in the Galaxy, our D.R.A.K.E. Equation estimates the number of animals in the known universe who may be detected via Zoom calls. The equation is as follows:
\begin{equation}
    N = R_Z \times n_h \times f_c \times f_{na} \times f_r \times f_p \times L ,
\end{equation}
with the following variables. Analogous to Drake's $R_*$ term, we adopt $R_Z$, the rate at which new Zoom calls are initiated. Our $n_h$ term is the average number of attendees per Zoom call. $f_c$ represents the fraction of Zoom call attendees who set their cameras to on. Other works such as \citet{allergies} make the point that we need to account for humans' being allergic to pets, so our $f_{na}$ term quantifies the percentage of humans who do not have a severe pet allergy. While many exopets will exist somewhere in homes, we can only detect those who make themselves visible. Our $f_r$ term estimates the fraction of a residence that a Zoom camera window captures. The $f_p$ term quantifies the fraction of exopets who are permitted by their hosts to enter the Zoom window area. Our final term, $L$, is the average length of a Zoom call.

While the D.R.A.K.E. equation is a very useful to organize one's thoughts and research about how many domestically residing animals and kitties are in the world, we acknowledge that making an accurate calculation is quite difficult due to large uncertainties in many of the involved terms. What follows is our rough order-of-magnitude estimation.

In April 2020, Zoom reported 300 million meeting participants daily (where one person attending $n$ meetings counts as $n$ participants)\footnote{Statistic from https://www.businessofapps.com/data/zoom-statistics/}.
In the authors' experience, most Zoom calls include at least 1 but fewer than 100 humans, so we estimate an average $n_h$ value of 10. We also use this number to estimate the number of Zoom calls opened per day, $R_Z$, as 10 million.
We adopt 0.5 for our $f_c$ value, fraction of zoom cameras on. This is roughly based on resonant participation field theory, which  accounts for the fact that in large Zoom calls, most users turn cameras off, while in small calls, most cameras are on (Astrobites, in prep). According to \citet{allergies}, 10-20\% of the human population is allergic to cats or dogs. So, we estimate our non-allergic fraction, $f_{na}$ to be 0.85. As an order of magnitude estimate, we adopt 0.1 for the fraction of a residence which is visible in a Zoom camera window, $f_r$. In the authors' experience, most humans allow their pets to enter the Zoom camera region, so we estimate $f_p$ as 1. Lastly, we adopt 1 hour as the standard Zoom call length, $L$. We encourage all parts of the community to adopt this as a standard as well.

So, our estimate for the number of exopets which are detectable via Zoom is:
\begin{equation}
    N = 10,000,000~\rm{days}^{-1} \times 10 \times 0.5 \times 0.85 \times 0.1 \times 1 \times 1~\rm{hour} \sim 10,000 .
\end{equation}
This immense number provides an optimistic outlook for the future of exopet detection science. We hope it provides a rallying call for the community to embrace their Zoom calls as an opportunity to conduct additional exopet science with no additional overhead. There are many more exciting detections to come. 

\section{Conclusions and Future Work} \label{sec:conc}
% YB quick summary
We present here observations of a new and intriguing paw-pulation of objects. These exopets are seen at wide size and mass ranges, and exist around a variety of hosts. Despite this variation, we have presented several observational parameters with which we can make inferences about their fundamental properties. 

Throughout this study, we have made a great deal of assumptions about the human hosts of these exopets, including that all hosts are the same size. This, of course, is unrealistic, so additional follow-up of the human hosts will be necessary to properly characterize the exopets and to eliminate false positives. % CC - I can expand on this and make it more fun if necessary

Additionally, all of the exopet detections discussed in this work have been made with instruments with optical wavelength ranges, but there is much more information to be found about these objects with multi-wavelength follow-up programs. In particular, we note that the surface temperature of these objects is expected to be $\sim$300 Kelvin, and so their blackbody spectrum should peak in the mid-infrared. This conveniently falls within the wavelength range of JWST (0.6-28 microns). We eagerly look towards the day that new exopets are identified with this flagship mission. %should be writing my paper, but I think sneaking in JWST (and who isn't these days) is just too good an opportunity to miss - MP
Moreover, following up exopet phenomena in the X-ray regime (especially for systems with high kinetic energy) will be plausible with future X-ray missions that are sensitive and appropriately named, e.g., Lynx \citep{lynx2018}. %love the JWST reference, and was wondering if a Lynx reference works too!! - GK 

\section*{acknowledgments}
This work is independent of the American Astronomical Society (AAS) and does not necessarily reflect the views of the entire organization. Thank you to everyone who contributed their pet sightings for this work. We love Astrobites! 


\vspace{5mm}


\software{NumPy \citep{numpy},
IPython \citep{ipython}, 
Jupyter Notebooks \citep{jupyter},
Matplotlib \citep{matplotlib}, 
Astropy \citep{astropy:2013, astropy:2018}, 
SciPy \citep{scipy}}


\newpage 
\appendix

\section{Definitions}

%----------------
%%%%A suggestion to add a definition table for key terms - GK

\begin{longtable}[c]{|c|c|}
\hline
\textbf{Term} & \textbf{Definition}\\ \hline
Floofiness                & Measurement of how extended the fluffy fur ``atmosphere'' of an exopet is \\ \hline
Snuggliness          &  Tendency of an exopet to cuddle with their hooman \\ \hline
SNAKEs          &  Sub-Neptunian Animal Keplerian Extended bodies   \\ \hline
DUCK Objects          &  Dynamically Unstable Coplanar Kepler Objects   \\ \hline
D.R.A.K.E. Equation          &  The Domesticly Residing Animal and Kitty Estimation Equation \\ \hline
\caption{Key terms and definitions used in this paper.}
\label{tab:definitions}\\
\end{longtable}

%----------------

\section{In-Situ Observations}\label{sec:insitu}

After collecting data on this new sample of exopets, we were able to send probe missions to multiple exopet systems in-situ in order to gather more detailed information about the nature of the observed phenomena. Although exoplanets are too far away for missions to currently reach, exopets are much closer and easier to observe. One of these exopet systems, the dog Briley b (Rocky), was observed multiple times in our sample and with direct imaging follow-up. Another system hosts two exopets: Catherine b and Catherine c, or Europa and Enceladus, respectively. Sabina b shows the snuggleness feature discussed in Section \ref{sec:transits}. While these exopets were not in our original sample, they provide a unique opportunity for in-depth study of both the ``cat'' category of exopets, as well as multi-pet systems. Furthermore, we show in-situ observations of the two clasess of exotic exoplanets discussed in Section ~\ref{sec:exotic}, namely Ben B and Mark B which are SNAKEs, as well as Olivia b, c, and d, a multi exopet system composed entirely of DUCKs.

\subsection{Briley b (a.k.a. ``Rocky'')}

\begin{figure}
    \centering
    \includegraphics[width=0.49\textwidth,angle=90]{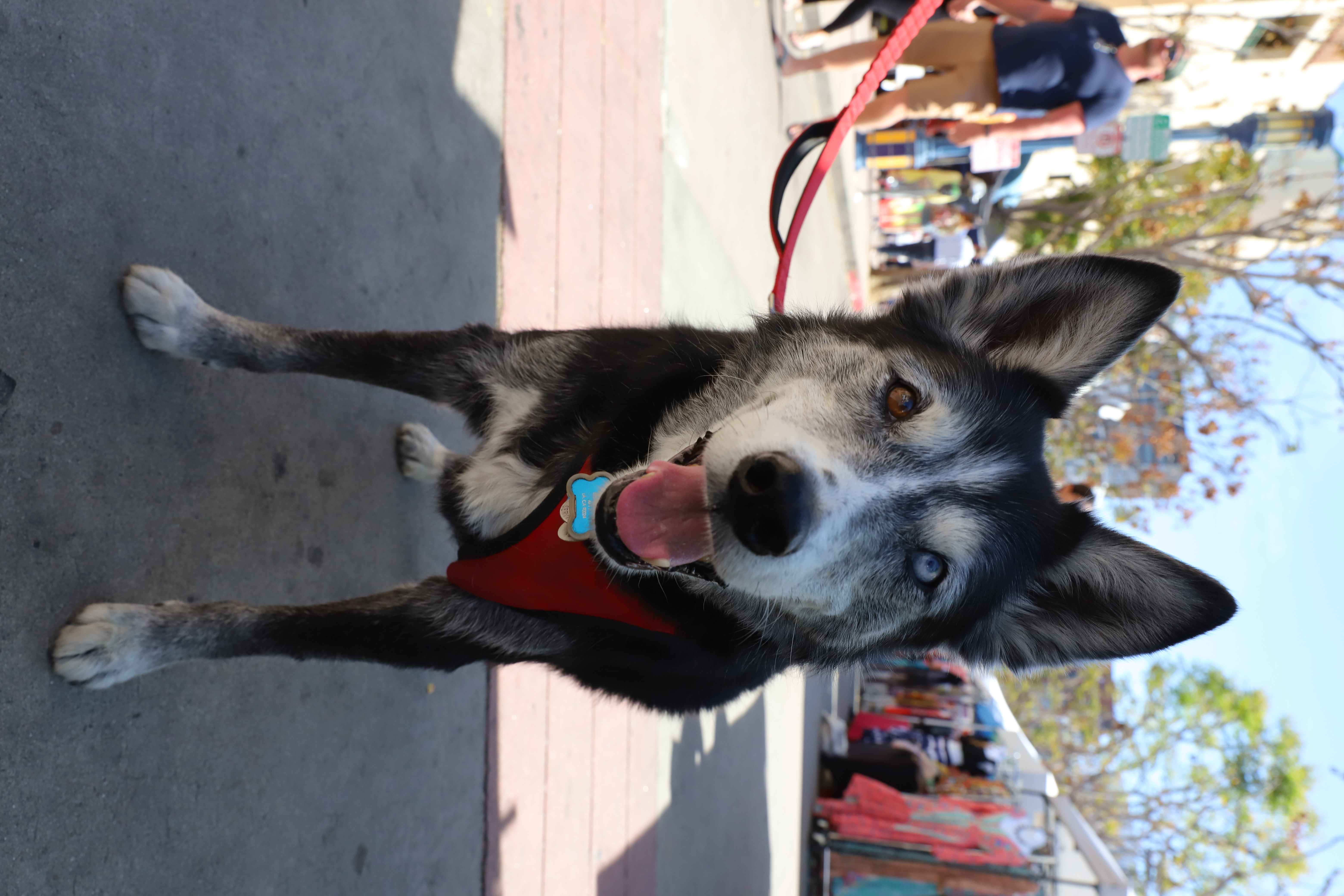}
    \includegraphics[width=0.49\textwidth,angle=90]{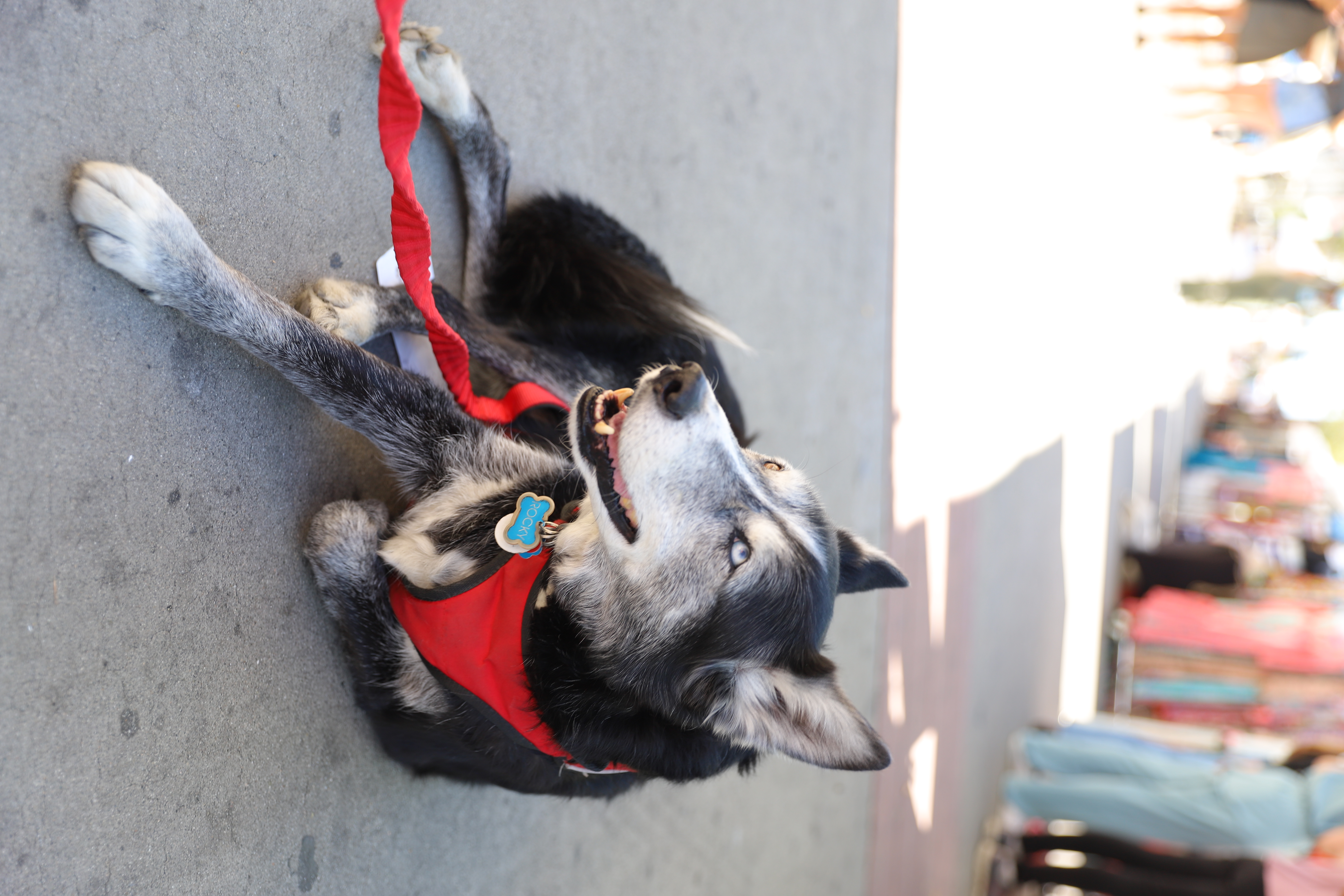}
    \includegraphics[width=0.49\textwidth,angle=90]{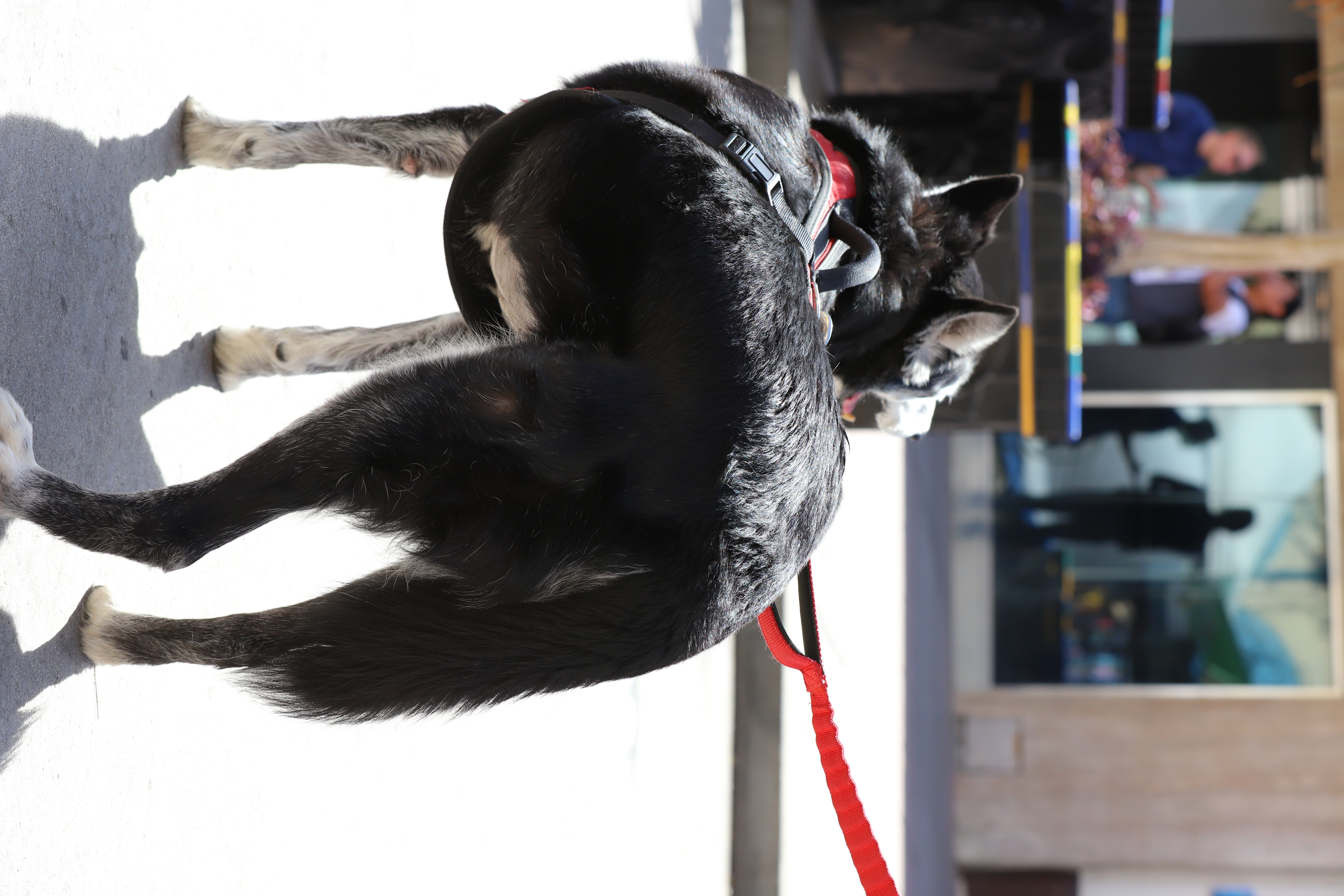}
    \includegraphics[width=0.49\textwidth]{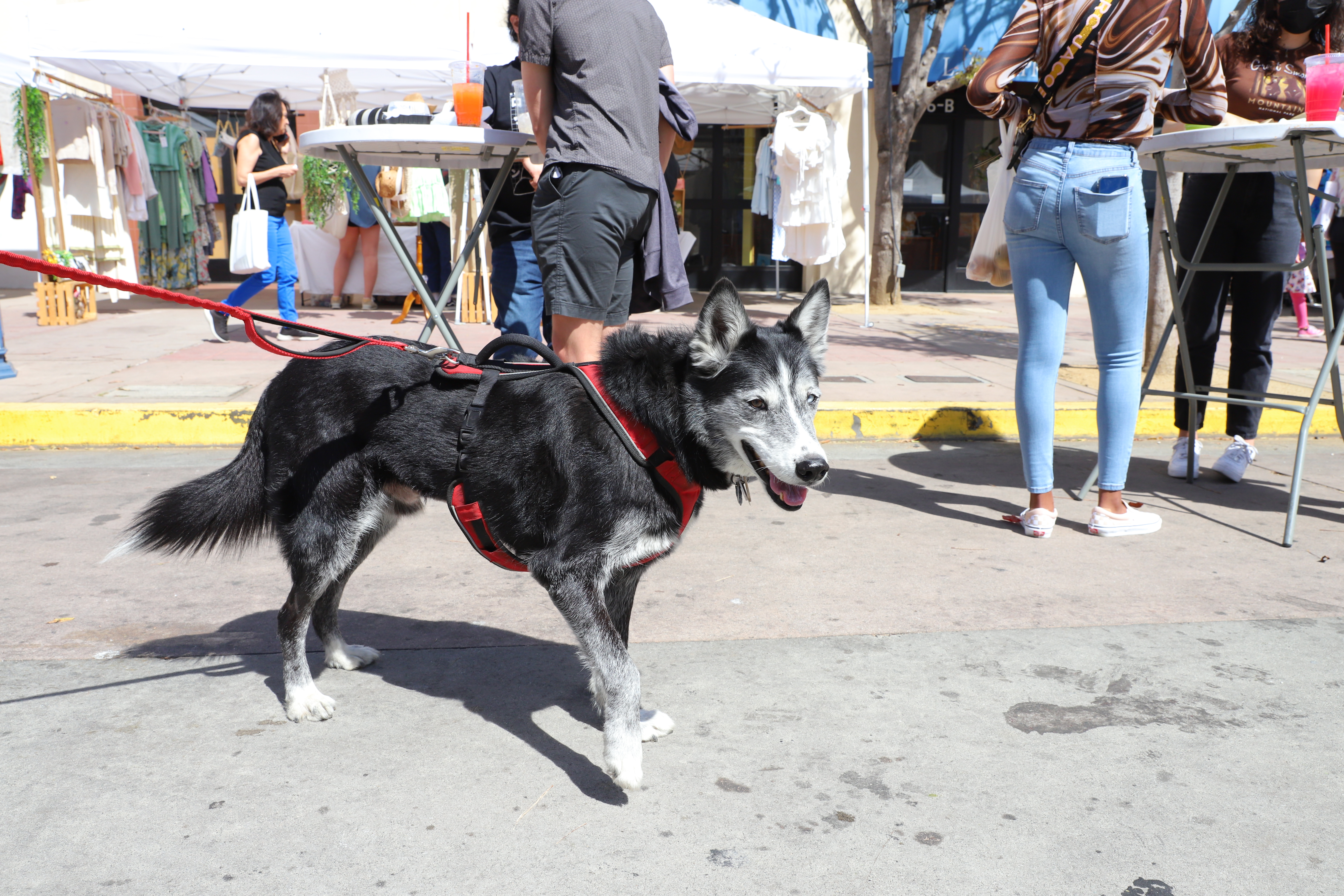}
    \caption{In-situ observations of Briley b, known colloquially as ``Rocky'', at different phase angles. These close-up observations reveal previously unknown details about exopets and challenge some of our original assumptions, indicating the need for more detailed models of floofiness, shape, color, and transit orientation. We were even able to capture the motion of this exopet, as shown in the bottom image. The images here were generously provided by observer Phil Travis of UCLA.}
    \label{fig:in_situ_dogs}
\end{figure}

In-situ observations confirm Rocky has a mass of approximately 65$\pm$5 pounds, and note that his radius is larger than approximated in transits, indicating that his orbit is further from his host star than originally thought. This may be typical of the dog variety of exopets, since they cannot transit as closely to their human hosts due to their lower ability to climb on tables. Rocky is fluffy, with particularly soft ears, and is black with detailed white and grey variations including very cute speckles on his snout. His floofiness varies across his body, with significantly more floofiness in his tail and neck, illustrating that floofiness need not be uniform across the entire exopet. For the first time, we were able to directly observe an exopet's motion with Rocky, and note that the motion of the exopet's limbs may affect their transit shape. Additionally, we note that the shape of this exopet is distinctly non-spherical in close-up observations, as shown in Figure \ref{fig:in_situ_dogs}, indicating that assuming spherical symmetry for the dog class of exopets may be inaccurate. Data indicates he is a 100/10 good boy. 

\subsection{Catherine b and c (a.k.a. ``Europa'' and `` Enceladus'')}

In-situ observations of Catherine b (a.k.a. ``Europa'') indicate that Europa has a mass of approximately $10\pm1$ pounds. Europa is fluffy and soft; we assign her floofiness a value of 3. Her coloring is unique; while she appears entirely black from afar (albedo $=0.09$), our in-situ observations reveal that she is in fact flecked with patches of orange. Additional spectroscopic measurements will be necessarily to fully characterize her compelling composition. In general, our in-situ observations of Europa indicate that she is best cat.

\begin{figure}
    \centering
    \includegraphics[width=0.49\textwidth]{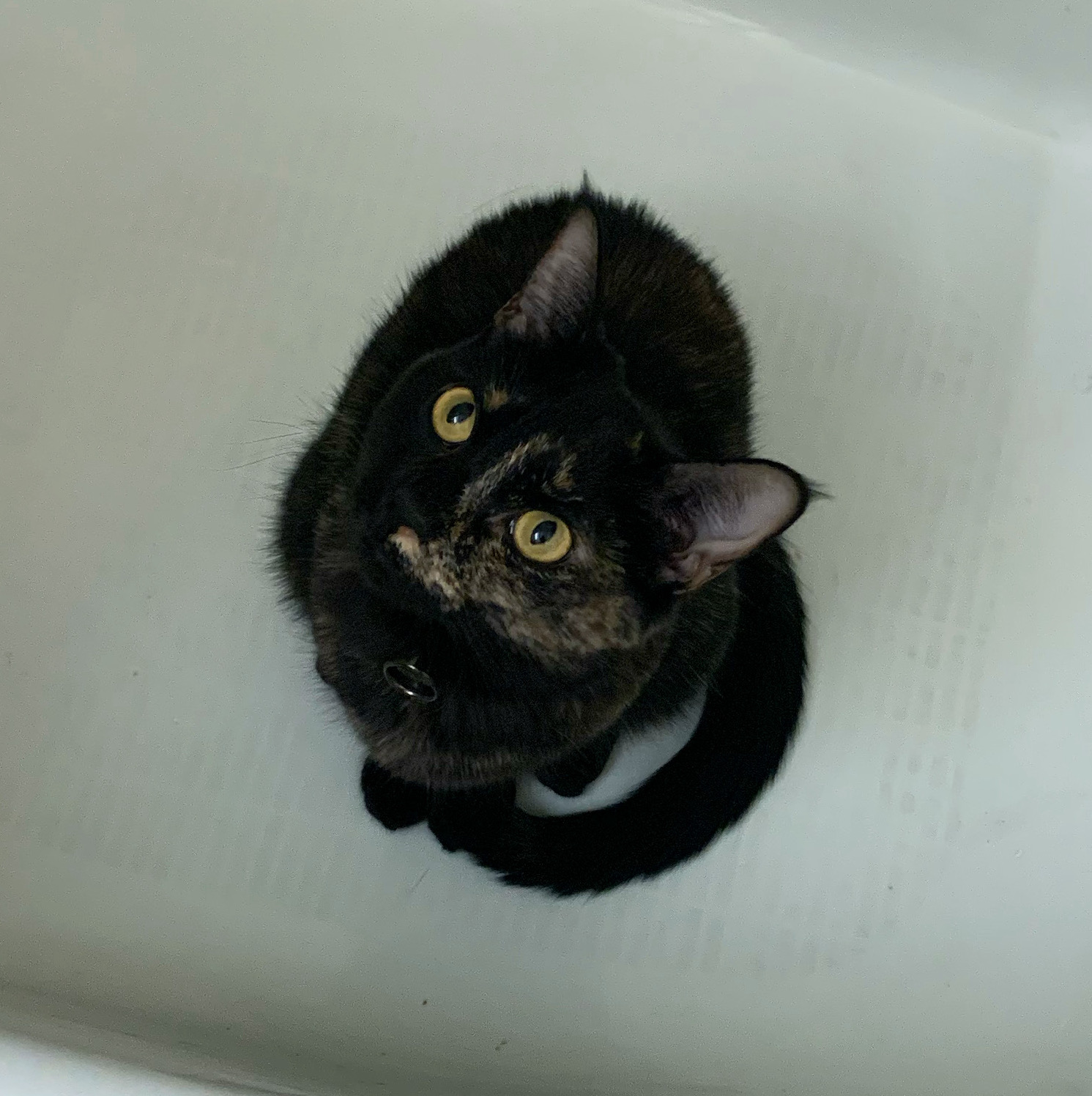}
    \includegraphics[width=0.49\textwidth]{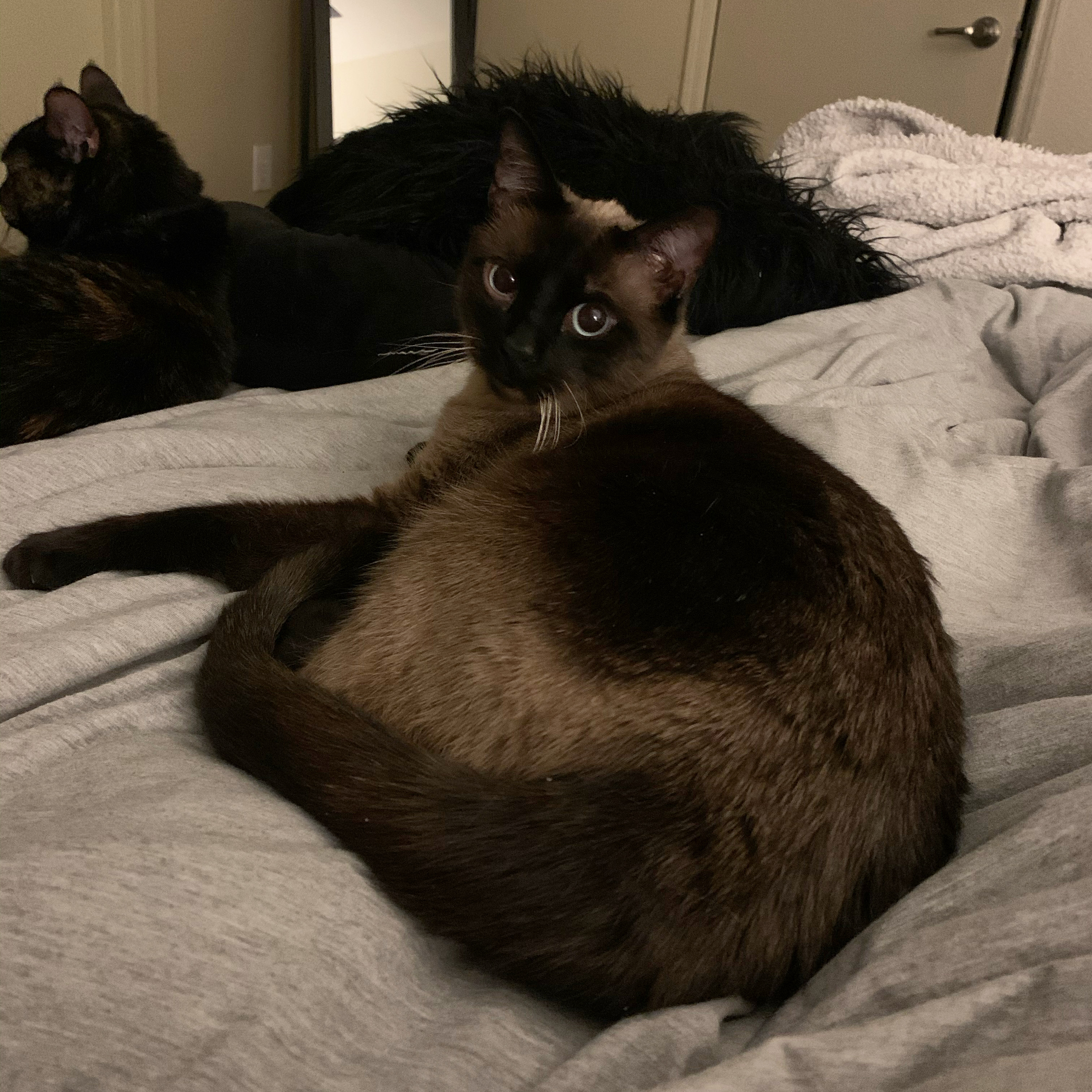}
    \caption{In-situ observations of Catherine b (left) and c (right), known colloquially as ``Europa'' and ``Enceladus'', respectively. We note that Europa is visible in the Enceladus observation as well. Though we do not use the spherical symmetry assumption proposed by \citet{Mayorga2021} in this work, we note that these exopets are quite spherical. These observations of Europa and Enceladus reveal that their coloring and thus their composition is more complex than originally thought, emphasizing the need for follow-up and in-situ measurements.}
    \label{fig:in_situ_cats}
\end{figure}

In-situ observations of Catherine c (a.k.a. ``Enceladus'') indicate that Enceladus has a mass of approximately $12\pm1$ pounds. Enceladus demonstrates similar softness to Europa; we also assign his floofiness a value of 3. His coloring is unique as well. He appears to be a ``Siamese'' subtype within the ``cat'' category of exopets; however, upon approaching Enceladus with our probe, we found that he is actually a mixture of multiple subtypes. This could indicate a unique formation mechanism involving a collision of two cat-sized exopets. Additional spectroscopic measurements, and perhaps a lander to probe the interior of Enceladus, will be needed to fully characterize his composition. Our observations indicate that Enceladus is also best cat. It may be that all cat exopets are best cat. Images from our in-situ observations of Europa and Enceladus are shown in Figure \ref{fig:in_situ_cats}. Note that Catherine b and c should not be confused with the Jovian or Saturnian moons.

\subsection{Sabina b (a.k.a. ``Pluto'')}

\begin{figure}
    \centering
    \includegraphics[width=0.49\textwidth, angle=270]{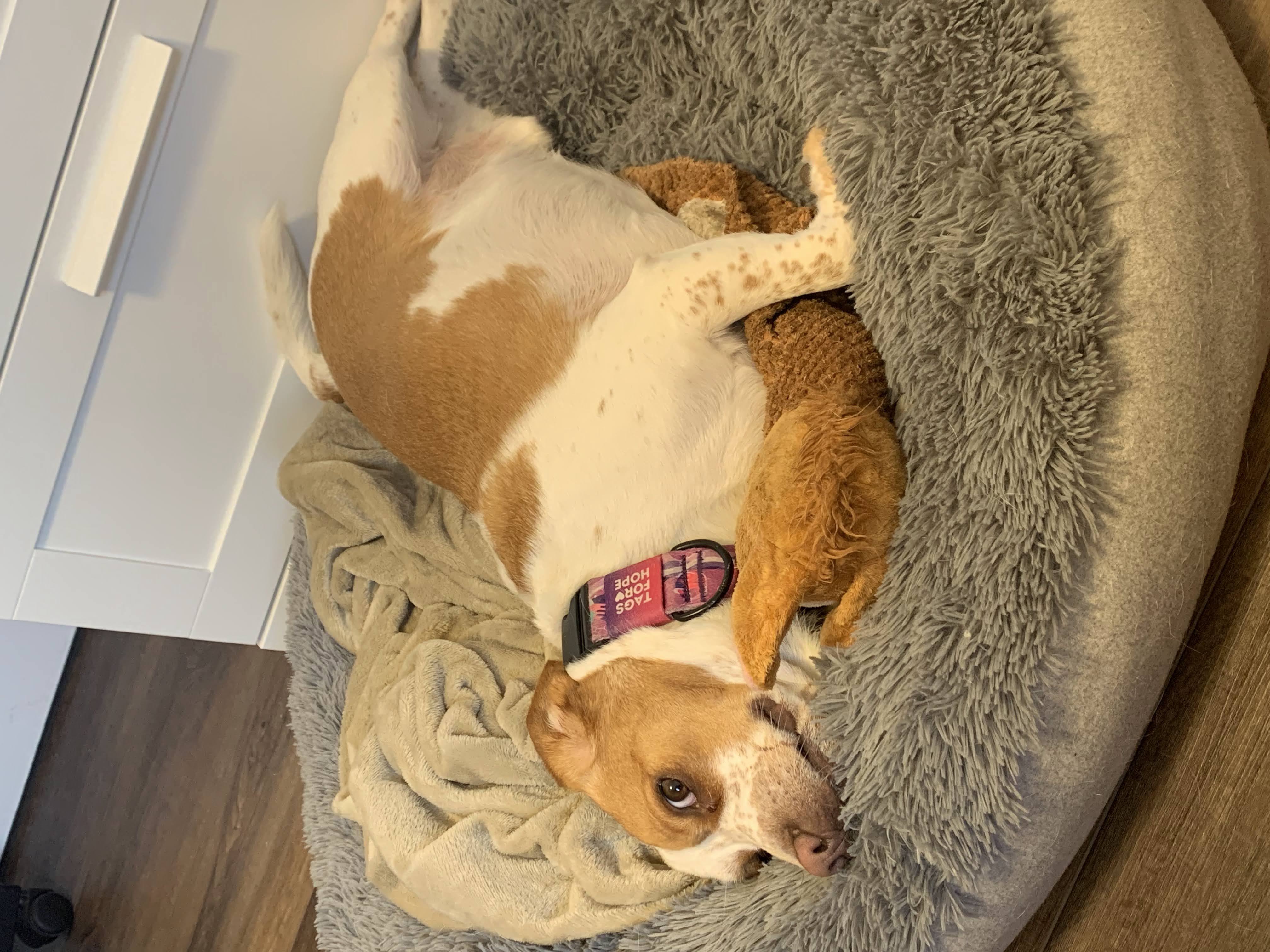}
    \includegraphics[width=0.49\textwidth, angle=270]{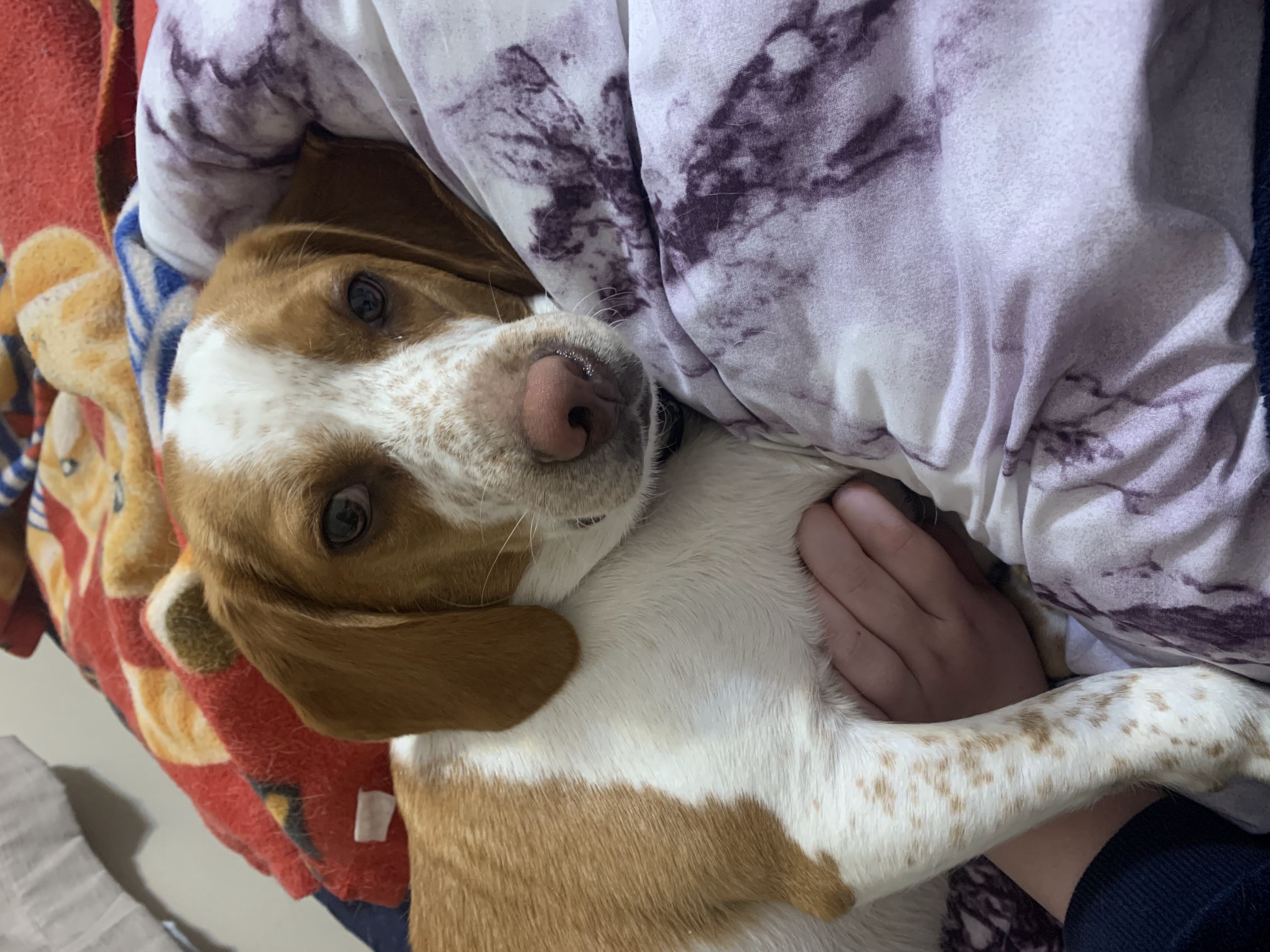}
    \caption{In-situ observations of Sabina b, known colloquially as ``Pluto.'' Left hand image shows the large neutron-star-like density mentioned above. Right hand image shows the high value of snuggliness also shown by our analysis.}
    \label{fig:in_situ_pluto}
\end{figure}

In-situ observations confirm Pluto has a mass of approximately 35$\pm$2 pounds. The first look at the exopet indicates that he is 50\% a beagle in terms of his breed; however, the other 50\% are not clear. The flux coming out of Pluto is very large due to his desire to snuggle, therefore we also conclude that he is a "lap dog." Some features on his surface, also known as freckles, and his love for chairs indicate that he used to be a prince but got cursed and transformed into an exo-dog. He is reminiscent of a neutron star in terms of his large density (Figure \ref{fig:in_situ_pluto}) and large rotational velocity whenever his host h\textit{oo}man (Sabina) interacts with food. Initial analysis also shows that his level of noise exceeds $10^6 dB$. In terms of good boiness, he is 100\% a good boy. 

\subsection{SNAKEs in the Mark and Ben Systems}

\begin{figure}[h]
    \centering
    \includegraphics[height=0.3\linewidth]{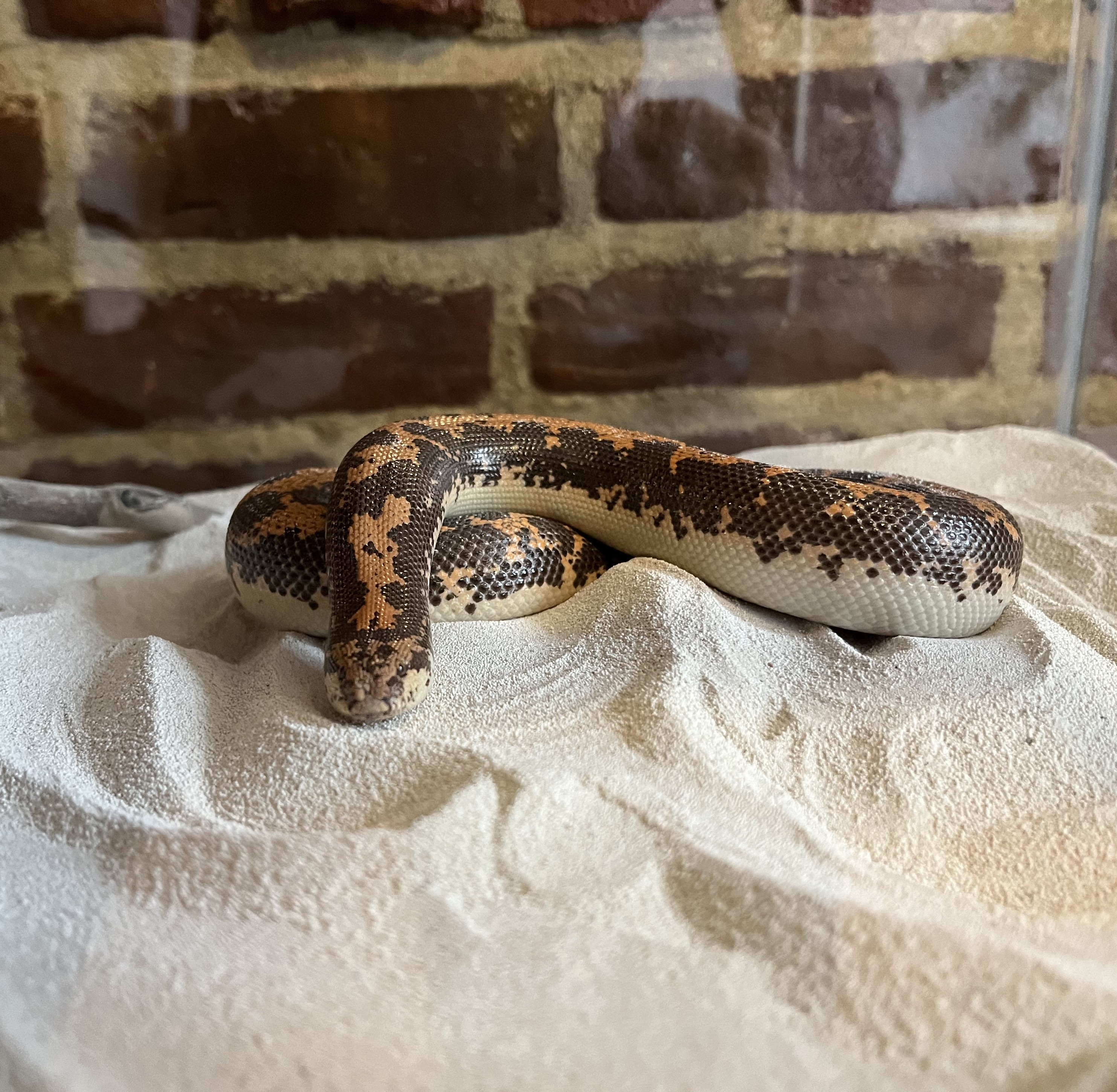}
    \includegraphics[height=0.3\linewidth]{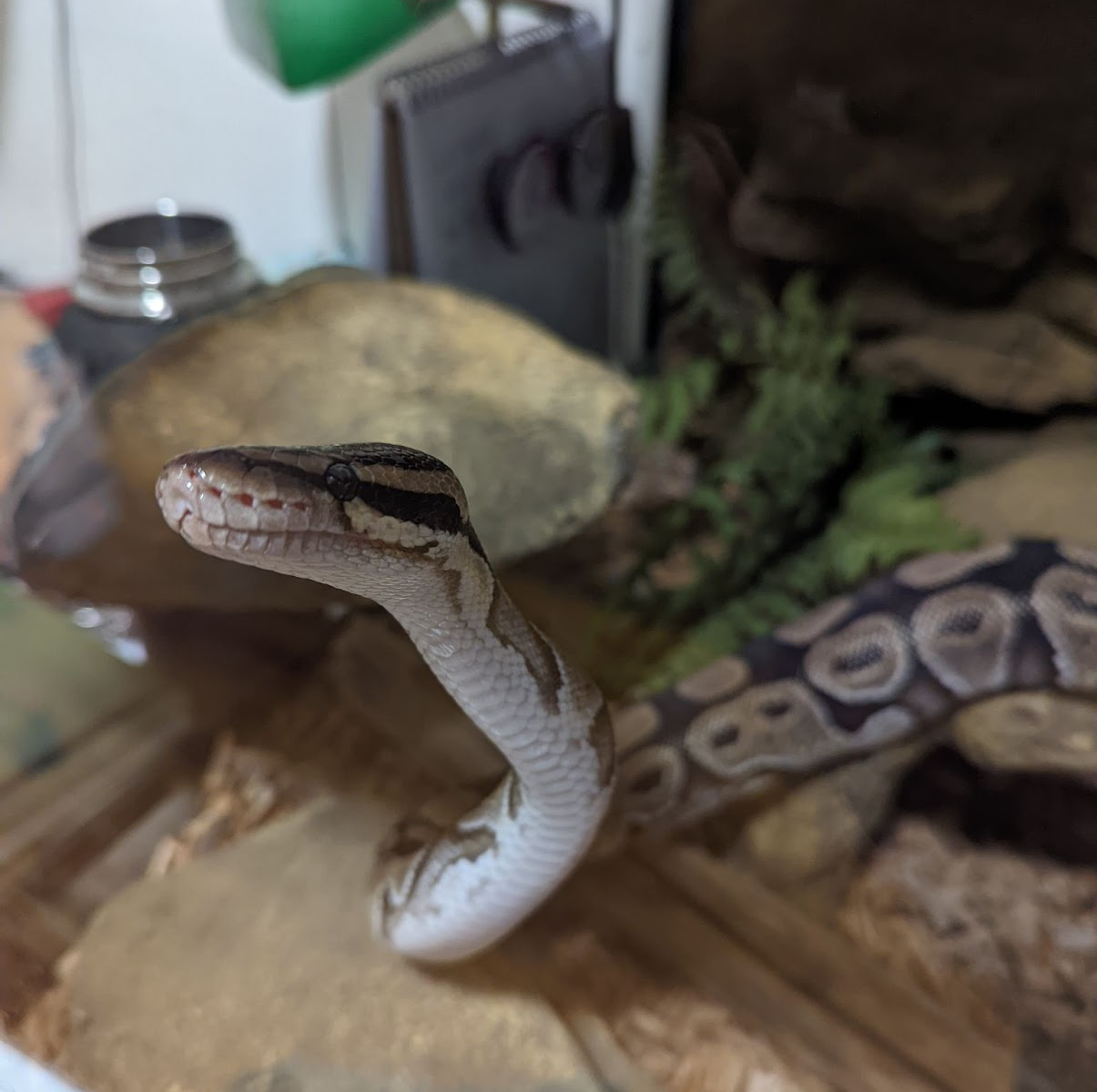}
    \caption{Targeted direct imaging observations of two SNAKEs. As noted, these targets were strongly polariz\st{ing}ed, a feature possibly responsible for their rarity in the backgrounds of Zoom calls. \textbf{Left:} Mark b (Sandy), classified as a Kenyan Sand Boa. \textbf{Right:} Ben b (Rutabaga), classified as a Ball Python.}\label{fig:snakes}
\end{figure}

In-situ observations of the SNAKEs sub-sample as shown in Figure \ref{fig:snakes} provide the first examples of this exotic exopet subclass. Models predict they have a slightly lower effective temperature, and a scaled surface rather than a floofy atmosphere. Undersampled in our Zoom based study, it is yet to be seen if this class of exopets is to be widely accepted by the community.

\subsection{DUCKs in the Olivia System (a.k.a. ``Brisket'', ``Chorizo'', \& ``Ceviche'')}

\begin{figure}
    \centering
    \includegraphics[width=0.53\textwidth, angle=0]{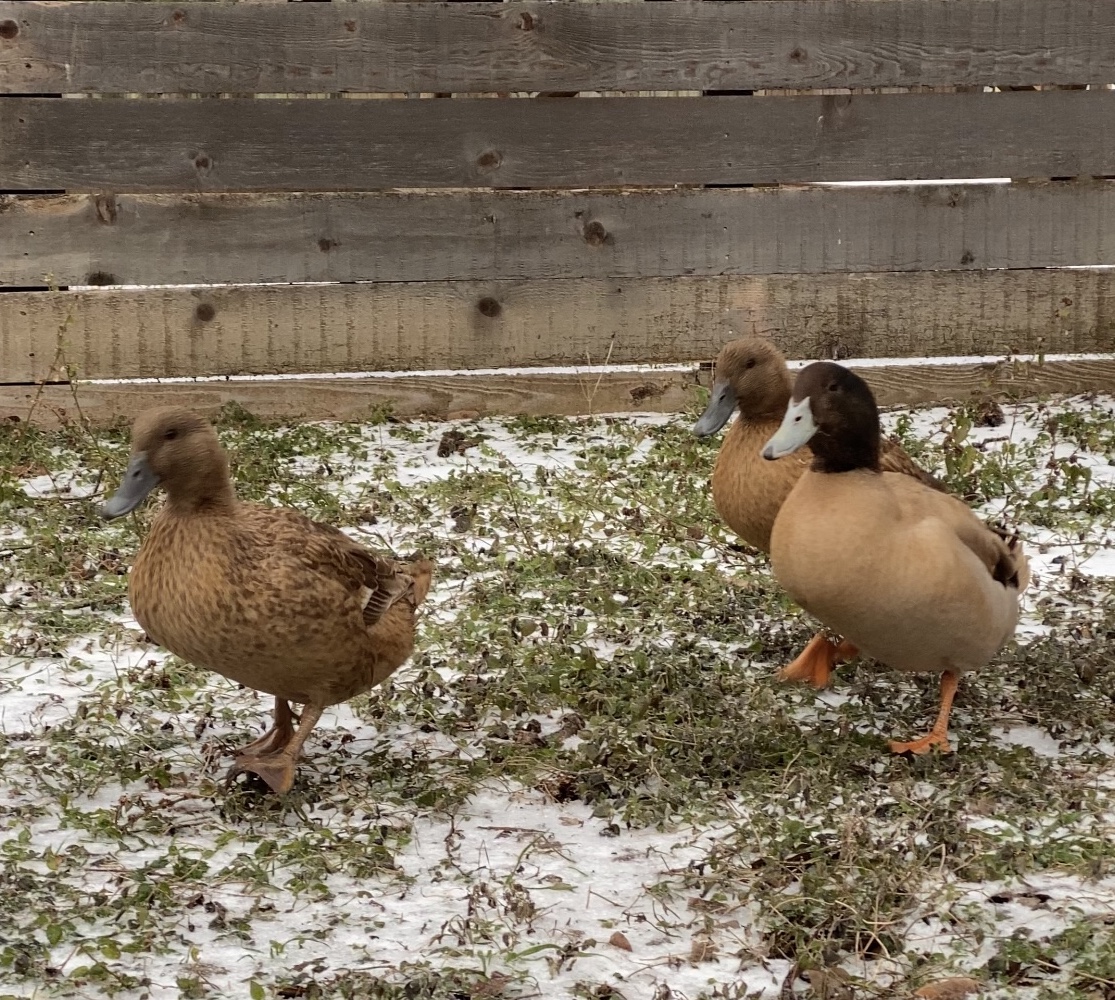}
    \caption{In-situ observations of a small group of DUCKs in a relatively stable orbit during a Cold DUCK Time. Their color is well characterized as Brown, while other properties remain poorly constrained. The imaging data shown here do not fully describe their dynamically unstable orbits.}
    \label{fig:in_situ_DUCKs}
\end{figure}

In-situ observations of the DUCK sub-sample as shown in Figure \ref{fig:in_situ_DUCKs} provide preliminary characterization of this exotic object type. Note that many properties of these objects are not well constrained, including the Floofiness and orbital parameters, given their unstable ``flying'' orbits and overall exotic nature within the exopet category. Future follow up is needed to more fully characterize this class of objects. 

\clearpage

\section{All Observational Data}\label{app-data}
\begin{center}\textit{See next page for data table.}\end{center}
\begin{longrotatetable}
\begin{deluxetable}{cccccccrr}
\tabletypesize{\tiny}
\tablecaption{Summary of Observational Data on Exop(lan)ets}

\tablehead{\colhead{Observer} & \colhead{Planet Designation} & \colhead{Planet Name} & \colhead{Event Type} & \colhead{Color} & \colhead{Floofiness} & \colhead{Animal Type} & \colhead{Transit depth} & \colhead{Transit Duration} \\ 
\colhead{} & \colhead{} & \colhead{(if known)} & \colhead{} & \colhead{} & \colhead{} & \colhead{} & \colhead{(\% of screen)} & \colhead{(sec)} } 

%% All data must appear between the \startdata and \enddata commands
\startdata
Abi & Abi b & Cordelia & Microlensing & Spotted/Mixed & 2 & Goat & 50 & 300 \\
Haley & Advisor b & Noel & Microlensing & Spotted/Mixed & 2 & Bird & 5 & 600 \\
Ali & AE b & Nova & Microlensing\tablenotemark{*} & Black & 5 & Cat & 35 & 300 \\
Florian & Anon b &  & Microlensing & Spotted/Mixed & 5 & Cat & 1 & 3 \\
Shirin & Anthony b &  & Transit & Brown & 3 & Dog & 10 & 10 \\
Avery & Avery b & Andromeda & Microlensing & Grey & 2 & Cat & 20 & 180 \\
Jillian & BE b & Calliope & Transit & Spotted/Mixed & 2 & Cat & 50 & 1800 \\
Brianna & Brianna b & Piper & Transit\tablenotemark{*} & Black & 3 & Dog & 10 & 10 \\
Briley & Briley b & Rocky & Transit & Black & 3 & Dog & 3 & 1200 \\
Briley & Briley b & Rocky & Transit\tablenotemark{*} & Black & 3 & Dog & 5 & 180 \\
Briley & Briley b & Rocky & Transit\tablenotemark{*} & Black & 3 & Dog & 15 & 240 \\
Briley & Briley b & Rocky & Transit & Black & 3 & Dog & 16 & 420 \\
Briley & BW b &  & Transit & Spotted/Mixed & 1 & Dog & 10 & 300 \\
Grace & Celina b & Salem & Microlensing & Black & 4 & Cat & 30 & 5 \\
Cristiano  & CH b &  & Transit & White & 5 & Cat & 20 & 30 \\
Melissa & CJ b &  & Transit & Grey & 3 & Cat & 50 & 3 \\
Arianna & CMC b & Franklin & Microlensing & Spotted/Mixed & 1 & Pig & 100 & 60 \\
Aman & CNL b &  & Transit & Grey & 4 & Cat & 36 & 300 \\
Tim & CO b &  & Transit & Grey & 3 & Cat & 80 & 10 \\
Taylor & CP b &  & Transit & White & 2 & Cat & 95 & 300 \\
Peter & Devon b & Charlie & Transit & Brown & 3 & Dog & 20 & 600 \\
Emily & Emily b & Peanut & Transit & Spotted/Mixed & 4 & Bird & 35 & 35 \\
Connor & ES b & Spammy & Transit & White & 3 & Cat & 40 & 120 \\
Jenny & Felipe b & Cheeto & Transit & Orange & 3 & Cat & 90 & 45 \\
Elizabeth & GR b &  & Transit & Spotted/Mixed & 2 & Cat & 5 & 15 \\
Jenny & Haley b & Newton & Transit & White & 4 & Dog & 10 & 120 \\
Sumeet & HD b & Jiggy & Transit & Brown & 2 & Dog & 20 & 3600 \\
Sasha & JB b &  & Transit & Grey & 2 & Cat & 2 & 1800 \\
Douglas & Jessica b & Monkey & Transit\tablenotemark{*} & Orange & 2 & Cat & 5 & 300 \\
Briley & JG b &  & Transit & Brown & 3 & Cat & 50 & 3 \\
Victor & JLarson b &  & Transit & Black & 3 & Cat & 70 & 120 \\
Ali & JN b & Tsaya & Microlensing & Grey & 1 & Cat & 25 & 5 \\
Suzi & JT b & Dusty & Microlensing & Black & 5 & Cat & 50 & 120 \\
Jenny & Kaitlin b & Mostly & Transit & Black & 2 & Cat & 15 & 120 \\
Mary & LB b & George & Transit & Grey & 3 & Cat & 10 & 300 \\
Leonardo & Lizvette b & Love & Transit & Spotted/Mixed & 2 & Cat & 26 & 300 \\
Rebecca & Lizvette b & Love & Transit & Brown & 3 & Cat & 33 & 10 \\
Macy & Macy b & Pepper & Microlensing\tablenotemark{*} & Black & 4 & Cat & 5 & 10 \\
Macy & Macy c & Charlie & Transit\tablenotemark{*} & Grey & 2 & Cat & 25 & 300 \\
Ali & Matthew b & Sylvia & Microlensing\tablenotemark{*} & Black & 2 & Cat & 15 & 30 \\
Sandip & MC b & Olive & Microlensing & Brown & 2 & Dog & 20 & 60 \\
Arianna & MFC b &  & Transit & Orange & 3 & Cat & 100 & 30 \\
 & Michael B & Luna & Imaging\tablenotemark{*} & Black & 2 & Dog & 100 & 60 \\
Peter & Mike Schreiner b &  & Transit & Grey & 3 & Dog & 10 & 300 \\
Jenny & Neil b &  & Transit & Orange & 5 & Dog & 15 & 120 \\
Anne & Niels b & Henk & Transit & Black & 4 & Bird & 5 & 2 \\
Macy & NT b & Phoenix & Microlensing\tablenotemark{*} & Spotted/Mixed & 2 & Cat & 10 & 20 \\
Sam & Olivia b & Nova & Transit & Brown & 2 & Dog & 1 & 5 \\
Briley & PD b & Hamilton & Microlensing & Black & 4 & Cat & 25 & 60 \\
Briley & PD b & Hamilton & Transit & Black & 4 & Cat & 15 & 240 \\
Briley & PD b & Hamilton & Transit & Brown & 5 & Cat & 24 & 300 \\
MJ & R b &  & Transit & Brown & 3 & Cat & 10 & 1200 \\
Briley & RB b &  & Transit & Spotted/Mixed & 2 & Cat & 5 & 90 \\
Briley & RB c &  & Transit & Spotted/Mixed & 2 & Cat & 4 & 120 \\
Briley & Reyna b &  & Transit & Black & 1 & Dog & 7 & 900 \\
Ryan & Ryan b & Arlo Betts Trainor & Transit\tablenotemark{*} & Black & 3 & Cat & 60 & 15 \\
Oli & S b &  & Transit & Brown & 4 & Cat & 20 & 5 \\
 & Sabina b & Pluto & Imaging\tablenotemark{*} & Ginger/White & 3 & Dog & 100 & 10 \\
Jessie & Sean b & Rusty & Transit\tablenotemark{*} & Orange & 3 & Cat & 60 & 5 \\
Peggy & TC b & Clyde & Transit & Spotted/Mixed & 4 & Cat & 40 & 90 \\
Ali & TE b & Schooner & Microlensing\tablenotemark{*} & Brown & 1 & Dog & 10 & 15 \\
Thomas & Thomas b & Callie & Microlensing\tablenotemark{*} & Spotted/Mixed & 1 & Cat & 15 & 30 \\
Kelle & Tia b &  & Transit & White & 4 & Dog & 20 & 10 \\
P & V b & M & Transit & Grey & 3 & Cat & 10 & 30 \\
\enddata

\tablecomments{Data from our exop(lan)et survey. Note objects Briley b (a.k.a. Rocky) and Lizvette b (a.k.a. Love) as systems with multiple observations of the same transiting object, and Macy b/c and RB b/c as the two multiple object systems in our sample.}
\tablenotetext{*}{Imaging data is also available for these observations.}
\end{deluxetable}
\end{longrotatetable}

\bibliography{main}
\bibliographystyle{aasjournal}



\end{document}